\documentclass{emulateapj}
\usepackage{xspace}
\usepackage{amsmath}
\usepackage{hyperref}
\usepackage{framed} 
\usepackage{txfonts}
\def\tsf{\tau_{\rm SF}}
\def\tbw{\tau_{\rm BW}}

\def\Msun{{\rm M}_\odot}

\def\ion#1#2{{\rm  #1}\textsc{#2}}
\def\HI{{\ion{H}{I}}}

\def\H2{{{\rm H}_2}}

\def\dr{{\Delta r}}
\def\dm{{M_1}}
\def\qcon{{\hat{Q}}}
\def\pmod{{\,{\cal P}}}

\def\kny{{k_{\rm Ny}}}

\def\bfull#1#2#3{B#1.#2.FULL.#3} 
\def\bgd14#1#2#3#4{B#1.#2.GD14#3.R#4} 

\def\dim#1{\mbox{\,#1}}
\def\hide#1{}

\begin{document}

\title{Cosmic Reionization On Computers: Numerical and Physical Convergence}

\author{Nickolay Y.\ Gnedin\altaffilmark{1,2,3}}
\altaffiltext{1}{Particle Astrophysics Center, Fermi National Accelerator Laboratory, Batavia, IL 60510, USA; gnedin@fnal.gov}
\altaffiltext{2}{Kavli Institute for Cosmological Physics, The University of Chicago, Chicago, IL 60637 USA;}
\altaffiltext{3}{Department of Astronomy \& Astrophysics, The
  University of Chicago, Chicago, IL 60637 USA} 

\begin{abstract}
In this paper I show that simulations of reionization performed under the Cosmic Reionization On Computers (CROC) project do converge in space and mass, albeit rather slowly. A fully converged solution (for a given star formation and feedback model) can be determined at a level of precision of about 20\%, but such a solution is useless in practice, since achieving it in production-grade simulations would require a large set of runs at various mass and spatial resolutions, and computational resources for such an undertaking are not yet readily available.

In order to make progress in the interim, I introduce a weak convergence correction factor in the star formation recipe, which allows one to approximate the fully converged solution with finite resolution simulations. The accuracy of weakly converged simulations approaches a comparable, $\sim20\%$ level of precision for star formation histories of individual galactic halos and other galactic properties that are directly related to star formation rates, like stellar masses and metallicities. Yet other properties of model galaxies, for example, their $\HI$ masses, are recovered in the weakly converged runs only within a factor of two.
\end{abstract}


\section{Introduction}
\label{sec:intro}

As the study of cosmic reionization enters a (nothing short of a true) Renaissance, with numerous observational probes (from JWST and 30-meter class telescopes to 21 cm experiments) just about to increase the volume and quality of observational data hundreds-fold, the demand on theory to remain on par with the forthcoming data forces a careful reexamination of the accuracy of the current theoretical modeling. 

At present, numerical simulations offer the most accurate and realistic theoretical models of reionization. Hence, evaluating the precision of modern simulations is an important and timely theoretical task, while we are all eagerly waiting for the flood of new data.

Results from computer simulations can be considered as solutions to actual physical equations only if they are numerically converged. In practice, however, it is often exceptionally difficult or even plainly impossible to reach what the EAGLE team \citep{sims:eagle} termed "strong" convergence - i.e., the effective independence of simulation results of spatial and mass resolution \citep[c.f.][for an incomplete list]{misc:sh03,sims:njo07,sims:aquarius,sims:mgk08,sims:owls,sims:illustris,sims:eagle}. Without a proper convergence study however, the results from a numerical simulation may remain suspicious to be simply numerical artifacts.

In galaxy formation modeling the situation is exacerbated by the fact that we can never formulate the problem to solve as a system of basic physical equations, and always have to rely on subgrid models for star formation and feedback. Any such model is necessarily scale-dependent, so a strong convergence limit may not exist or be physically meaningless \citep[see][for a more elaborated discussion]{sims:eagle}.

A practical, although not necessarily mathematically rigorous solution to this dilemma is to seek "weak" convergence\footnote{I am again following the \citet{sims:eagle} terminology here.}, i.e.\ reduced dependence of the simulation results on spatial and mass resolution, when the actual numerical model to be simulated is adjusted as the resolution changes. This is precisely the goal I have in this paper: sufficiently accurate weakly converged simulations will serve as a robust theoretical model for the interpretation of the future observational data, and as the diversity, quality and volume of the data increase, the weakly converged numerical models are likely to improve further with the continuing effort.

In order to present the subject rigorously, I start with the discussion of strong and weak convergence in \S \ref{sec:conv}. As I have mentioned above, the actual  terminology has been introduced by \citet{sims:eagle}, so this presentation is not original, but it serves as a useful basis for the further exposition. In \S \ref{sec:sims} I enumerate original simulations used in this paper. All of the existing simulation work is united under a single project, Cosmic Reionization On Computers (CROC), that has been described in \citet{ng:g14}. Section \ref{sec:sfh} presents both my methodology and its application to achieving numerical convergence in the global star formation history of the universe for CROC simulations. Finally, \S \ref{sec:weak} describes the weakly converged numerical models that will serve as a physical basis for the next installment of CROC simulations.

\section{Numerical and Physical Convergence}
\label{sec:conv}

\subsection{Strong and Weak Numerical Convergence}
\label{sec:numconv}

Generically, any physical quantity extracted from a simulation needs to be tested for numerical convergence - a value far from convergence is meaningless, as it is dominated by numerical truncation errors. In the specific case of cosmological simulations, the two most crucial numerical parameters are spatial and mass resolution. The former is often fixed in either comoving or physical coordinates. In this paper I primarily consider simulations whose spatial resolution $\dr$ is fixed in physical coordinates, and the details of the implementation of the fixed physical resolution in the ART code are presented in the Appendix (\S \ref{app:resol}).

The mass resolution is convenient to parametrize by a quantity $\dm$, which is equal to the particle mass in a pure N-body simulation with the same number of particles, i.e
\[
  \dm \equiv \frac{\Omega_{\rm M}}{\Omega_{\rm M}-\Omega_{\rm B}} M_{\rm DM},
\]
where $M_{\rm DM}$ is the dark matter particle mass in the simulation. For example, a simulation with $20h^{-1}\dim{Mpc}$ box size and $1024^3$ particles would have $\dm = 9.2\times10^5\Msun$ for our adopted values of cosmological parameters ($\Omega_{\rm M}=0.3036$, $\Omega_{\rm B}=0.0479$, $h=0.6814$ in a flat $\Lambda$CDM cosmological model).

In principle, the third important numerical parameter is the dynamic range of initial conditions. For example, one could imagine a simulation with $1024^3$ particles but with initial conditions set up on a $512^3$ grid. In practice, however, overwhelming majority of all numerical studies in cosmology would use initial conditions on a $1024^3$ grid for a $1024^3$ particle simulation, and all CROC simulations are set the same way too. Hence, the parameter $\dm$ quantifies not only the mass resolution of the simulation, but also the dynamic range of initial conditions.

Hence, any physical quantity $Q$ extracted from the simulation (it can be literally anything, from a very local property like the density value in a given location to a global value, like the average star formation rate density) can be considered to be a function of numerical parameters,
\[
  Q = Q(\dr,\dm).
\]
If the cosmological simulations were produced from the ``first principles'', that would be the complete dependence. However, while many physical processes are modeled from the first principles in CROC simulations, star formation and stellar feedback are followed with phenomenological models; different star formation models\footnote{Hereafter, for the sake of brevity, I will use the term ``star formation model'' to mean both the star formation recipe and the model for the stellar feedback.} may result in different simulation predictions, and, hence, simulated quantities also depend on the physical model (I label it $\pmod$) and its parameters $p_j$,
\begin{equation}
  Q = Q(\dr,\dm | \pmod, p_j).
  \label{eq:qdep}
\end{equation}

Ideally, the fully numerically converged value $\qcon$ would be obtained as a limit of Equation (\ref{eq:qdep}) for infinitely fine spatial and mass resolution, but with physical model parameter fixed,
\begin{equation}
  \qcon = \lim_{\dr,\dm\rightarrow0} Q(\dr,\dm | \pmod, p_j),
  \label{eq:sconv}
\end{equation}
where hereafter I use a hat symbol $\hat{}$ to label a numerically converged result (notice, that so far it is not specified how the double limit is actually taken; I will discuss this later). Such a limit is commonly called ``strong convergence'' after \citet{sims:eagle}, in a sense that a physical model is fully independent of numerical resolution. Alas, as was mentioned in the Introduction and discussed by  \citet{sims:eagle}, in practice this is rarely achievable or useful, as phenomenological models typically are only valid for a finite range of spatial and mass scales. I.e., even if the limit of Equation (\ref{eq:sconv}) exists, it may make little physical sense, as the physical model $\pmod$ would not be valid for very small $\dr$. For example, an empirical linear Kennicutt-Schmidt relation used as a star formation recipe for CROC simulations is known to work well on scales of several hundred parsecs  \citep{ism:lwbb08,sfr:blwb08,sfr:gtgs10,sfr:dbwd10,sfr:bljo11,sfr:blwb11,sfr:lbbb12,sfr:lwss13,sfr:tngc13}, but fails or changes on scales of tens of parsecs \citep{ism:okt10,ism:sck14,ism:ehv14,ism:chc15}.

An alternative to strong convergence is ``weak convergence'', when parameters of the physical model are adjusted as the resolution increases,
\begin{equation}
  \qcon = \lim_{\dr,\dm\rightarrow0} Q(\dr,\dm | \pmod, p_j = f_J(\dr,\dm)).
  \label{eq:wconv}
\end{equation}
Weak convergence is less appealing, as the physical model is ``tweaked'' each time resolution changes, but the ultimate test is whether the converged values of parameters, 
\[
  \hat{p}_j = \lim_{\dr,\dm\rightarrow0} f_J(\dr,\dm),
\]
are physically meaningful. If they are, then there is nothing wrong with weak convergence. In fact, the EAGLE team \citep{sims:eagle} argued that weak convergence is physically more meaningful than formal strong convergence, since any subgrid model is phenomenological, and, hence, must be scale-dependent.

An alternative interpretation of weak convergence is that converged values $\hat{p}_j$ are the ``correct'' ones, but at finite resolution ($\dr,\dm$) parameters may be tweaked to produce a better converged value, i.e. $\qcon$ may be close to the unconverged value $Q(\dr,\dm | \pmod, p_j)$ for some values of $p_j$ even if the finite resolution value at the converged parameter values, $Q(\dr,\dm | \pmod, \hat{p}_j)$, is sufficiently far from the converged value $\qcon$.

\subsection{Physical Convergence}
\label{sec:physconv}

Just by itself, numerical convergence does not imply the truth. A numerically converged value may be wrong, if the underlying physical model is wrong. Hence, one should also consider ``physical convergence'' of a simulation, i.e. the sensitivity of $\qcon$ to the adopted physical model $\pmod$. Of course, it is impossible to consider any sort of a limit for $\pmod$, but one can explore the difference between simulation results with two or more different physical models,
\[
  \delta\qcon = \lim_{\dr,\dm\rightarrow0}\left[ Q(\dr,\dm | \pmod_1) - Q(\dr,\dm | \pmod_2)\right].
\]
If several sufficiently different physical models give similar results, it seems reasonable to assign higher confidence in the simulated values.

The physical model used in CROC simulations is fully described in \citet{ng:g14}. That paper also explored in sufficient detail the dependence of the star formation model on some of its parameters, most importantly the molecular hydrogen depletion time $\tsf$ and the feedback model delayed cooling time $\tbw$. There is a degeneracy between these two parameters (faster star formation can be compensated by stronger feedback), but the remaining parametric freedom can only be constrained by observational measurements. 

In this paper I do not explore the variation of these two parameters any further, considering the previous study sufficient for now. Hence, all of the results presented below are also subject to variation of $\tsf$, $\tbw$, and any other model parameter not explicitly discussed here.

An important component of the star formation model that has not yet been explored fully is the model for the formation of molecular hydrogen. In CROC simulations stars are only allowed to form in molecular gas, hence modeling the molecular gas is the cornerstone of the adopted star formation model. ART supports two methods for modeling molecular hydrogen - (A) a highly computationally expensive, full non-equilibrium chemical network, including $\H2$ formation on dust, and detailed radiative transfer of ionizing and ultra-violet radiation with accurate account for $\H2$ self-shielding \citep{ng:gk10,ng:gd14}, and (B) a simplified $\H2$ formation model based on fitting formulas from \citet{ng:gd14}\footnote{The fitting formulas are obtained by running a wide grid of galaxy formation simulations with the full non-equilibrium network. } and radiative transfer of ionizing radiation only. I will call the first physical model ``Full Chemistry'' and the second, simplified model ``GD14 Fits''. It is the second, simplified model that has been so far used in all CROC simulations, as it would be impractical to use the ``Full Chemistry'' model in large-scale, non-zoom-in simulations.

There exists, however, one serious limitation of the GD14 fitting formulas: they were calibrated on lower redshift ($z\sim3$) simulations, and do not account for the fact that molecular hydrogen formation time may be long, even longer than the age of the universe. At $z\sim3$ this is not an issue, since the age of the universe is $2\dim{Gyr}$, comparable to the molecular hydrogen depletion time. At high redshifts, $z\ga10$, the effect of the finite age of the universe is, however, substantial; in particular, at $z>10$ original GD14 fits would produce unphysical global star formation rate in excess of ``Full Chemistry'' simulations. Hence, I introduce a correction to the GD14 model, calling it ``GD14+''. This correction is defined and validated in Appendix \ref{app:h2mod}.

\section{Description of the Simulations}
\label{sec:sims}

A complete description of the physics followed in CROC simulations is presented in \citet{ng:g14} and I do not repeat it here for the sake of brevity.

\begin{table}[t]
\caption{Mass Resolution Chart\label{tab:mr}}
\centering
\begin{tabular}{llll}
\hline\hline\\
Label & $\dm$  & $N$ particles in                  & $N$ particles in \\
        &            & $10h^{-1}\dim{Mpc}$ box & $20h^{-1}\dim{Mpc}$ box \\
\\
\hline\\
``LR'' & $5.9\times10^7\Msun$  & $128^3$  &  \\
``MR'' & $7.4\times10^6\Msun$  & $256^3$  & $512^3$  \\
``HR'' & $9.2\times10^5\Msun$  & $512^3$  & $1024^3$ \\
``UR'' & $1.2\times10^5\Msun$  & $1024^3$ &  \\
\\
\hline\\
\end{tabular}
\end{table}

For exploring the convergence properties of CROC simulations, I use $10h^{-1}\dim{Mpc}$ and $20h^{-1}\dim{Mpc}$ boxes, as larger simulations would be too expensive for a sufficiently dense sampling of numerical parameters. Simulations that use GD14 fitting formulas (i.e., prototypes for production simulations) are run at fixed spatial resolutions, with values $\dr=25$, $50$, $100$, $200$, and $400\dim{pc}$, except for the ``ultra-high'' (UR) resolution, which would be impractical to sample fully with 5 spatial resolutions; only the lowest resolution, $\dr=400\dim{pc}$, is performed for this mass resolution. This is further discussed in \S\ \ref{sec:testm}.

Because enforcing constant resolution in physical units involves substantial amount of spatial smoothing, and, hence, resolution loss, reference simulations using the ``Full Chemistry'' model are run at constant comoving resolution.

Individual simulations are labeled in the following way: a simulation label starts with B10 or B20 for the box size and then is followed by a two-letter identification of the simulation mass resolution as shown in Table \ref{tab:mr}. The third field indicates the physical model used (``FULL'' for ``Full Chemistry'', ``GD14'' for GD14 fits, etc). The fourth field gives the spatial resolution in parsecs (c.f.\ ``R100'' stands for $\dr=100\dim{pc}$), if the resolution is fixed in physical units, or the maximum level of refinement (c.f.\ ``L7'' for $L_{\rm MAX}=7$) for simulations with constant comoving resolution. For reference, a simulation with $10h^{-1}\dim{Mpc}$, high mass resolution ($512^3$ root grid) and $L_{\rm MAX}=7$ has spatial resolution in physical units of $\dr = 20(11/(1+z))\dim{pc}$.

For example, \bfull{10}{HR}{L7} labels a simulation in a $10h^{-1}\dim{Mpc}$ box with $512^3$ particles, the ``Full Chemistry'' model, and spatial resolution fixed in comoving units, with $L_{\rm MAX}=7$, while \bgd14{10}{MR}{+}{200} labels a simulation in a $10h^{-1}\dim{Mpc}$ box with $256^3$ particles, the ``GD14+'' physical model (\S \ref{app:h2mod}), and fixed physical spatial resolution of $200\dim{pc}$. 

Simulations with progressively higher resolution but the same box size (like B10.LR, B10.MR, B10.HR, etc) start with initial conditions that preserve the same large-scale modes, so that the same individual objects can be identified in simulations with different mass resolution. I.e., initial conditions for, say, B10.MR are obtained from the initial conditions for B10.LR by adding small scale power between $\kny({\rm LR})$ and $\kny({\rm MR}) = 2\times\kny({\rm LR})$, initial conditions for B10.HR are further refined by adding small scale power between $\kny({\rm MR})$ and $\kny({\rm HR}) = 2\times\kny({\rm MR})$, etc, where $\kny(...)$ is the Nyquist frequency of the corresponding simulation box.

\section{Converging on the Global Star Formation History}
\label{sec:sfh}

One of the most generic quantities in a cosmological simulation is the globally averaged star formation rate density as a function of cosmic time \citep{misc:sh03,sims:svbw10}, often called ``global star formation history''. If a numerical simulation is not converged on the globally averaged star formation rate density, little else would be converged too. I also use this simple quantity to demonstrate in detail the methodology used for exploring convergence.

\subsection{Spatial Convergence at Fixed Mass Resolution}
\label{sec:testx}

\begin{figure}[t]
\includegraphics[width=\hsize]{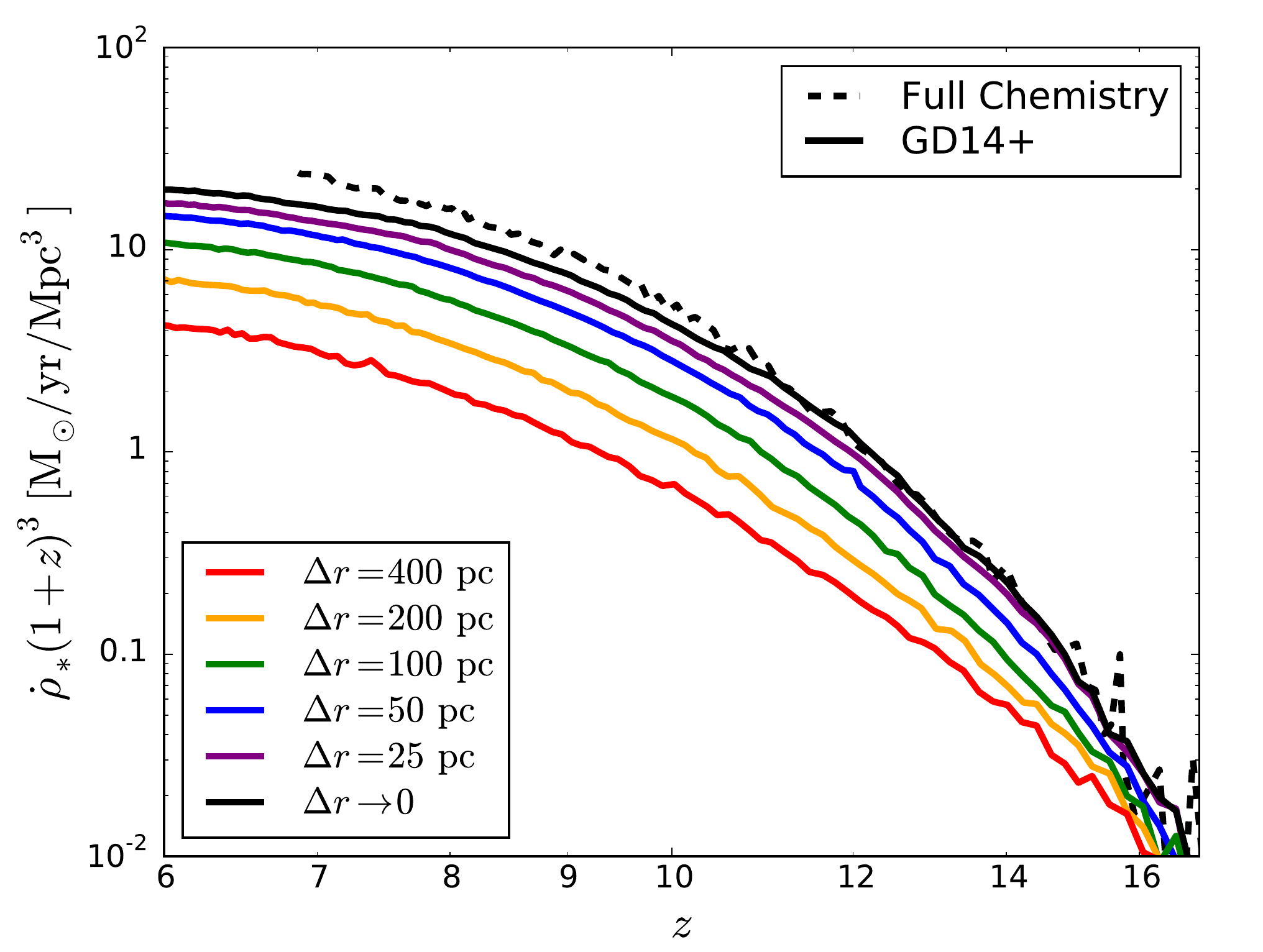}
\caption{Global star formation histories for several ``HR'' ($\dm=9.2\times10^5\Msun$) simulations. Various colored solid lines (from red to purple in rainbow order) show actual simulation results with spatial resolutions of $\dr=400$, $200$, $100$, $50\dim{pc}$, and $25\dim{pc}$ respectively for our default ``GD14+'' model, while a solid black line gives the numerically converged history. As a test of physical convergence, the dashed black line shows the fully converged reference ``Full Chemistry'' HR run. \label{fig:sfrxhr}}
\end{figure}

Figure \ref{fig:sfrxhr} shows such a convergence study for CROC high resolution ($\dm=9.2\times10^5\Msun$) simulations. Let's first focus on colored lines. They show $10h^{-1}\dim{Mpc}$ box simulations at several spatial resolutions (from $\dr=25\dim{pc}$ to $\dr=400\dim{pc}$) for the default ``GD14+'' model. Simulation results differ for different spatial resolutions, so even at the resolution of $\dr=25\dim{pc}$ simulations have not yet fully converged. Is it possible to extrapolate simulation results to find the converged answer?

As the default extrapolation scheme, I adopt a Taylor series expansion of the log of a quantity (in this case - the global star formation rate density at a given redshift),
\begin{equation}
  Q(\dr) = Q_0 \exp\left(-\sum_{i=1}^n C_j \dr^j\right),
  \label{eq:fitexp}
\end{equation}
where $Q_0$ is the spatially converged value,
\[
  Q_0 = \lim_{\dr\rightarrow0}Q(\dr).
\]
With 5 numerically sampled values of $Q$, one can determine $Q_0$ and up to first 4 coefficients $C_j$ ($n=4$). In practice, however, since simulation results are always somewhat noisy, it is preferably to use a smaller number of degrees of freedom than the number of data points, to avoid fitting numerical noise.

Since the extrapolating procedure does depend on the adopted function form for the fitting function, I also consider an alternative, 3-parameter power-law functional form,
\begin{equation}
  Q(\dr) = Q_0 \exp\left(-A \dr^B\right),
  \label{eq:fitpow}
\end{equation}
where $Q_0$, $A$ and $B>0$ are fitting parameters. I also experimented with other fitting functions, including using Taylor series expansion for the value itself, rather than its log ($Q(\dr) = Q_0 -\sum_{i=1}^n C_j \dr^j$), but these two functional forms given above seem to capture most of variations due to adopted parametrization.

\begin{figure}[t]
\includegraphics[width=\hsize]{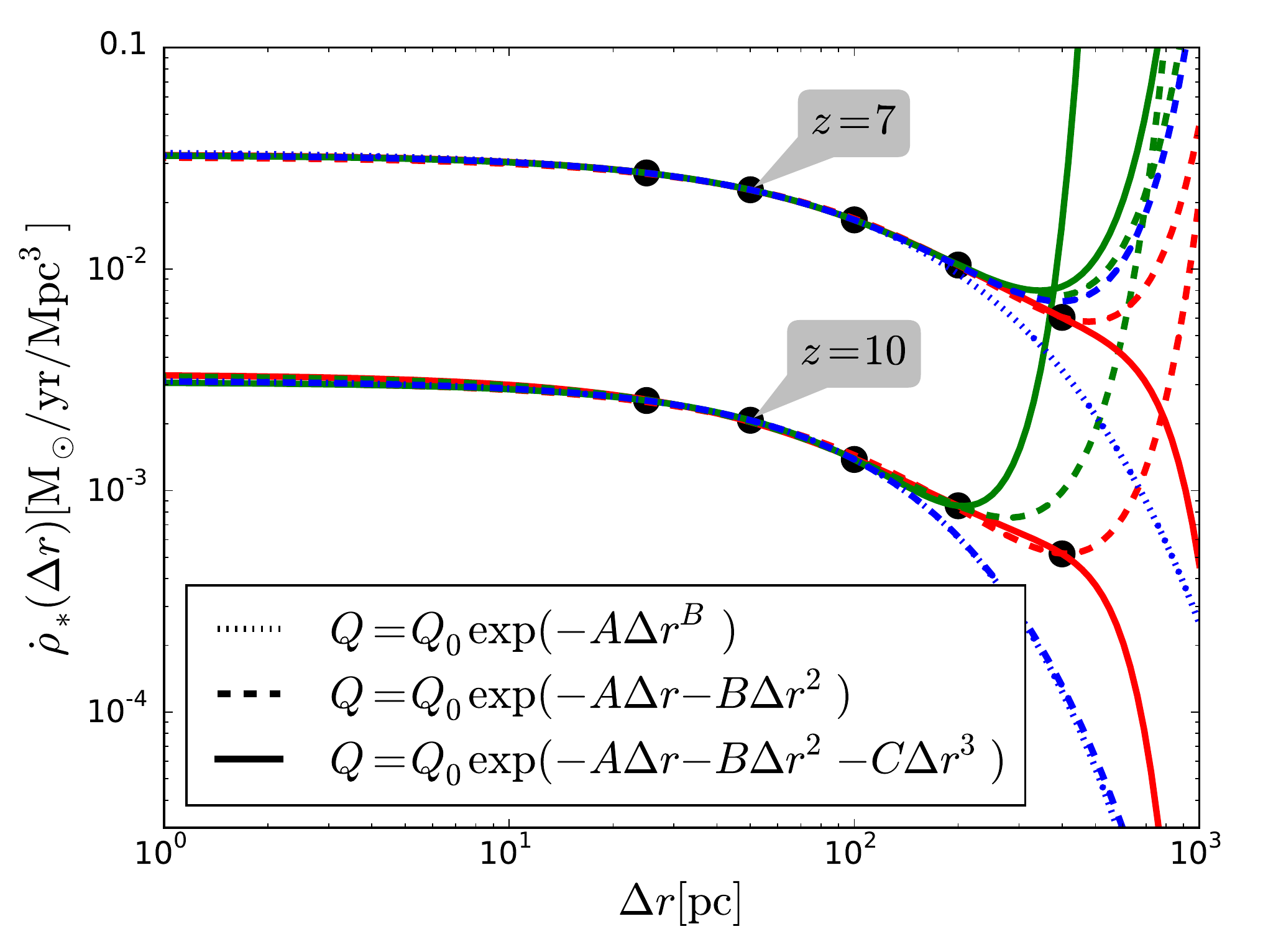}
\includegraphics[width=\hsize]{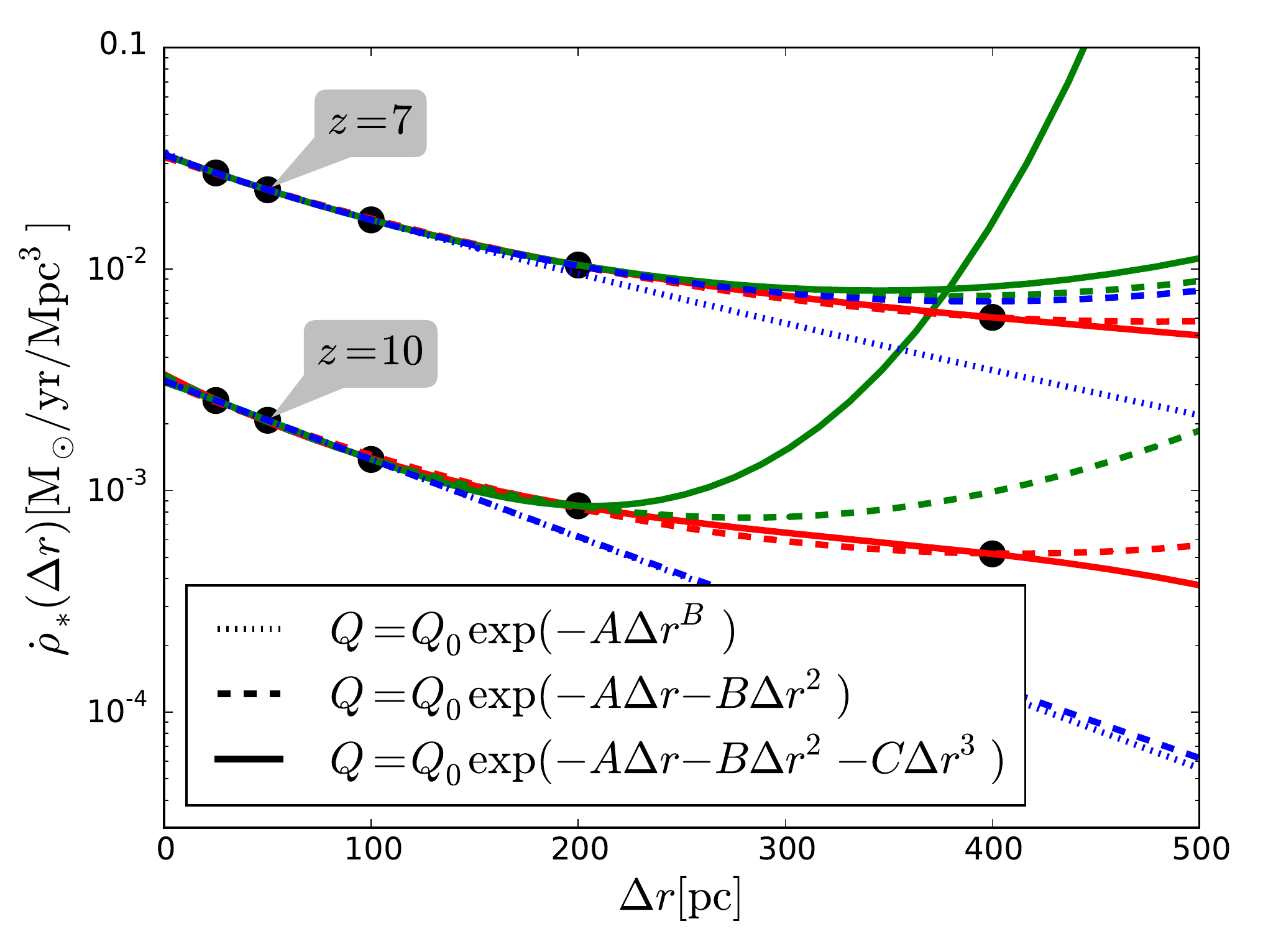}
\caption{Extrapolations of the simulated results at $z=7$ and $z=10$ to the numerically converged values along the spatial resolution axis. Black circles trace numerical results at the particular redshift, while lines show various extrapolations (see text for more details). Colors (red, green, and blue) correspond to cases when 5, 4, and 3 highest resolution data points are used for extrapolation. A numerically converged result (extrapolation to $\dr\rightarrow0$) can be determined to about 10\% precision. The two panels are the same figure with logarithmic and linear horizontal axes (the log-scaled plot is easier to see, but the actual amount of interpolation to $\dr\rightarrow0$ required is clearer in the linear space).\label{fig:tfitx}}
\end{figure}

As an example of the extrapolation technique, Figure \ref{fig:tfitx} shows cuts through Fig.\ \ref{fig:sfrxhr} at two representative redshift values, and results of extrapolating numerically sampled values with functions (\ref{eq:fitexp}) and (\ref{eq:fitpow}), using either all 5 data points or subsets of 4 or even 3 highest resolution points. Differences between these various choices serve as an estimate of the uncertainties in the extrapolation procedure, which in this case does not exceed 10\%.

Using the functional form (\ref{eq:fitexp}) with $n=3$ ($n=2$ for the ``Full Chemistry model) as the fiducial one, I can now extrapolate simulated results to the limit $\dr\rightarrow0$ at each redshift, and these extrapolations are shown with black lines in Fig.\ \ref{fig:sfrxhr}. Physical convergence can now be tested by comparing the fully converged  ``Full Chemistry'' simulations (the black dashed line in Fig.\ \ref{fig:sfrxhr}) and ``GD14+'' runs (a black solid line). The (level of) agreement between these three lines demonstrates the physical convergence of the CROC simulations; it is not, of course, achieved serendipitously, but simply the outcome of choosing the right values for the model parameters, $D_c=0.05$, in Equation (\ref{eq:qmod}). The agreement is very good at high redshifts, but deteriorates to about 20\% after $z\approx10$. One cannot say a priori which of the two solutions is the ``right'' one; while the ``Full Chemistry'' model includes more physics, it also takes the ``blastwave'' feedback model into the resolution limit where it is not supposed to work well. Hence, the difference between the black lines in Fig.\ \ref{fig:sfrxhr} should, at present, be treated as an estimate of the theoretical error due to the assumed $\H2$ formation model (and there must be other theoretical errors on top of this due to other physical assumptions).

The interpretation of this figure is as follows. It demonstrates that simulations with a finite spatial resolution $\dr>0$ do have a convergence limit $\dr\rightarrow0$; the limit is slightly dependent on the adopted extrapolating function, but is robust to within about 10\%, which can be treated as a ``systematic error'' of the numerically converged result. This numerically converged limit can be tuned to be within the 20\% variation from the "Full Chemistry" model, thus allowing one to use a simplified, "GD14+" model for molecular hydrogen formation as a reasonable approximation to a full non-equilibrium calculation.

\subsection{Mass Convergence at Fixed Spatial Resolution}
\label{sec:testm}

\begin{figure}[t]
\includegraphics[width=\hsize]{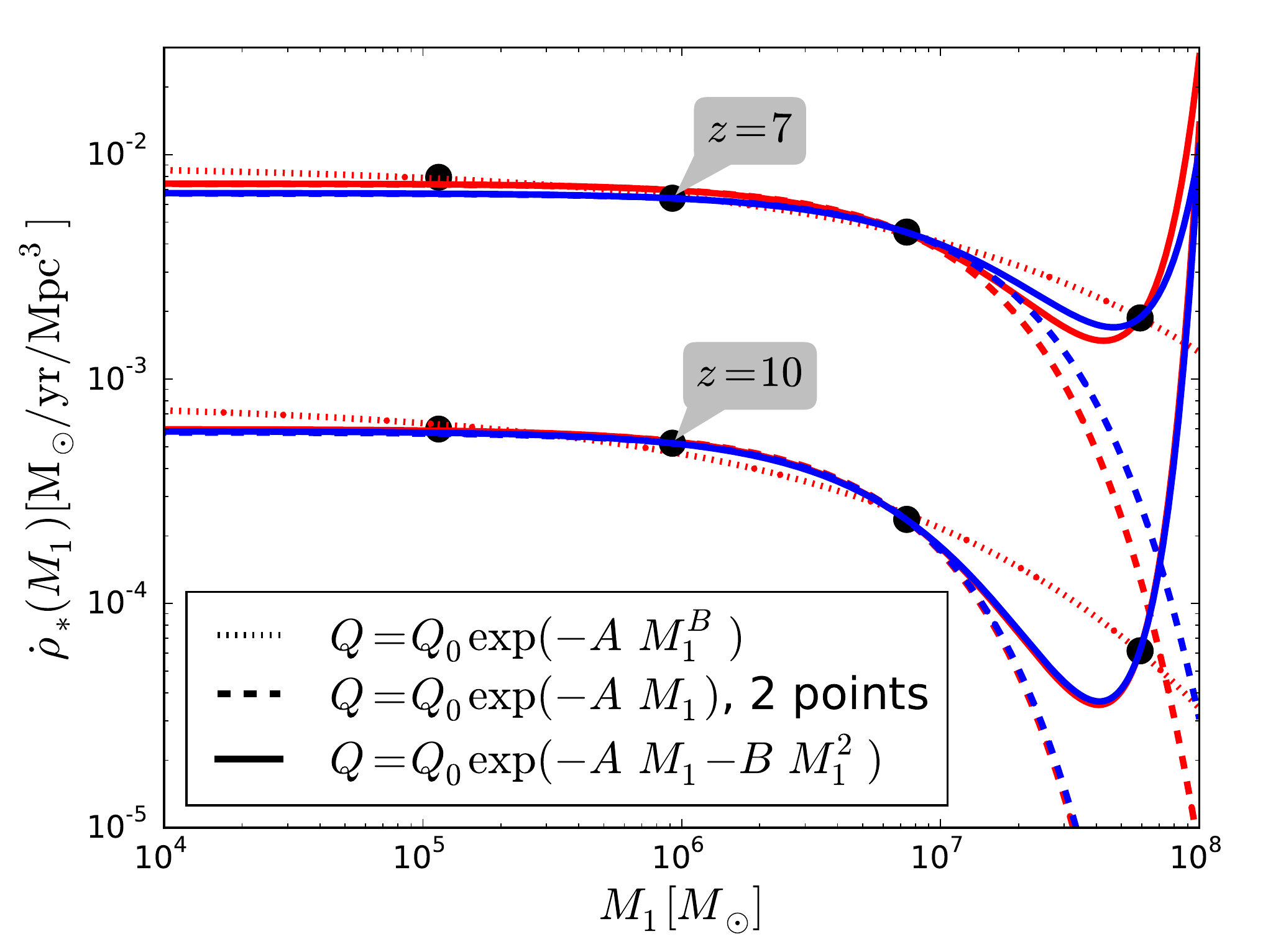}
\includegraphics[width=\hsize]{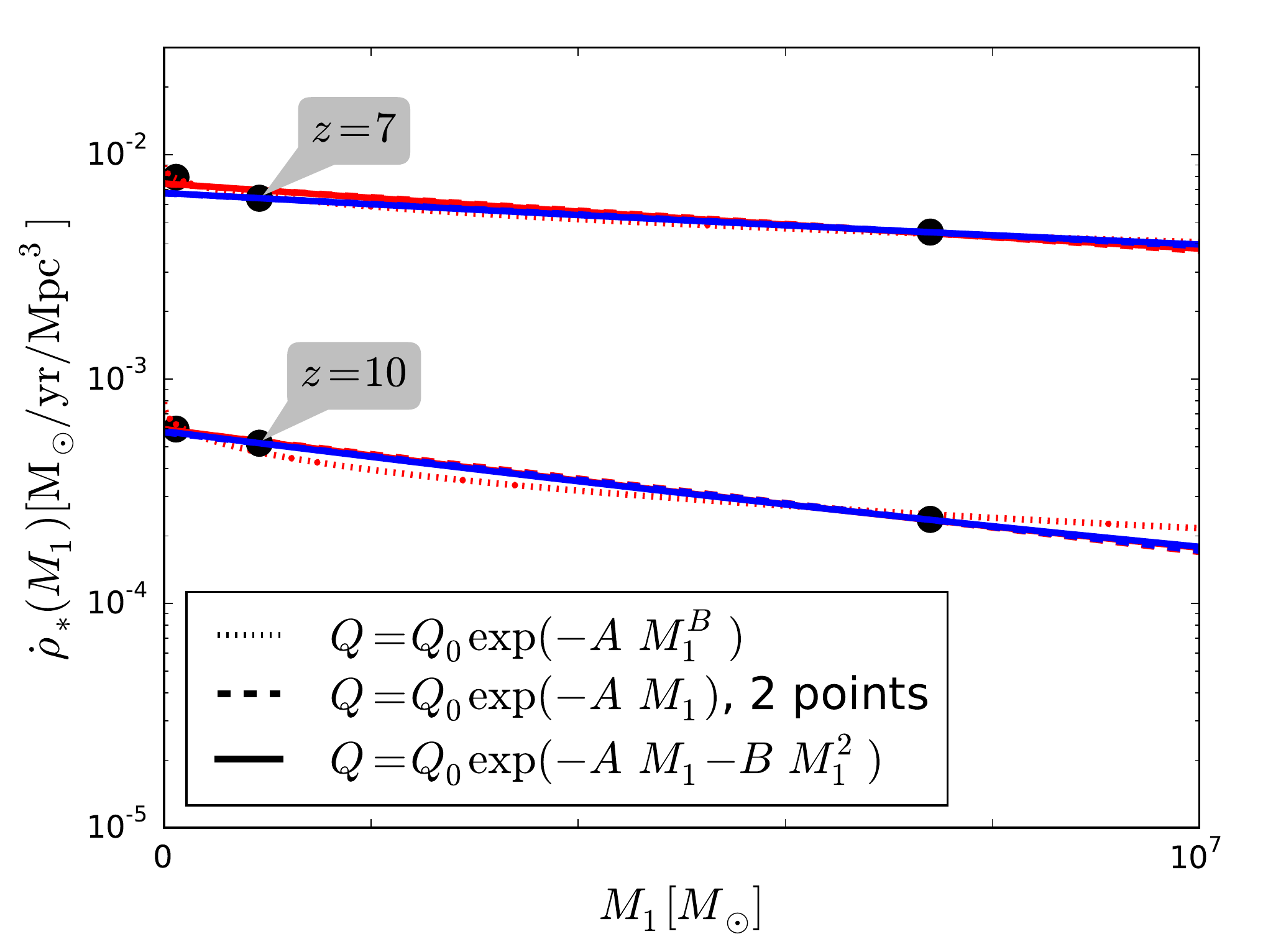}
\caption{Extrapolations of the simulated results at $z=7$ and $z=10$ to the numerically converged values along the mass resolution axis. Black circles trace numerical results at the particular redshift, while lines show various extrapolations (see text for more details). Colors (red and blue) correspond to cases when 4 and 3 highest resolution data points are used for extrapolation. A numerically converged result (extrapolation to $\dm\rightarrow0$) can be determined to about 20\% precision. The two panels are the same figure with logarithmic and linear horizontal axes (the log-scaled plot is easier to see, but the actual amount of interpolation to $\dm\rightarrow0$ required is clearer in the linear space).\label{fig:tfitm}}
\end{figure}

A similar procedure can be used to explore mass convergence. For a technical reason, the ART code requires the size of the root grid to be a power of two, so the mass resolution can, at present, be only sampled in factors of 8. That limits the number of samples we can have in most cases to just 3 (LR, MR, and HR). It is hard to estimate the robustness of the extrapolation procedure with just 3 samples, so one additional simulation with ``ultra-high'' (UR) resolution ($\dm=1.2\times10^5\Msun$) has been completed. At such an extreme mass resolution, it is only practical to run at the lowest spatial resolution, $\dr=400\dim{pc}$. Hence, a small corner of the full resolution sample grid ($\dm=1.2\times10^5\Msun$, $\dr\leq200\dim{pc}$) remains unexplored in this work. It can be filled in the future, as computing power increases even further.

Figure \ref{fig:tfitm} serves as a direct analog of Fig.\ \ref{fig:tfitx}, but now showing the extrapolation along the mass direction. Because the number of data samples is less than in the spatial extrapolation case, the variation between various extrapolating functional forms is greater. In the mass direction the converged value is only good to about 20\%, but that is sufficient at present, since the physical convergence is only good to at most that level.

\subsection{Full Numerical Convergence}
\label{sec:testf}

\begin{figure}[t]
\includegraphics[width=\hsize]{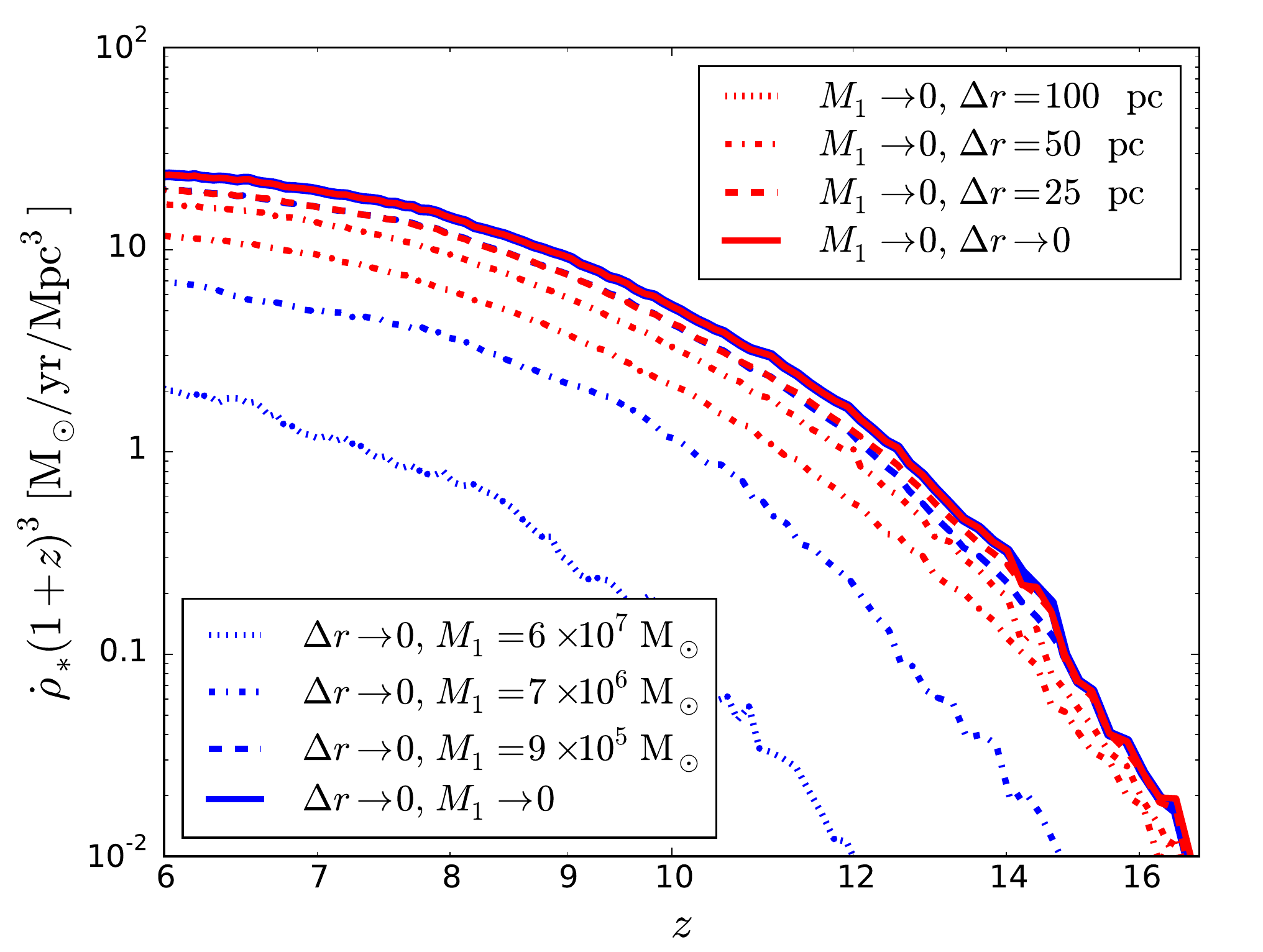}
\caption{Fully (i.e.\ spatially and mass) converged global star formation histories for CROC simulations. Red lines show the sequence of mass-converged values, non-solid lines tracking three fixed spatial resolution (out of total 5), while the red solid line is the limit $\dm\rightarrow0,\dr\rightarrow0$. Blue lines show the opposite approach, with the blue solid line tracing the limit $\dr\rightarrow0,\dm\rightarrow0$. Solid blue and red lines coincide, demonstrating the stability of extrapolation.\label{fig:sfrall}}
\end{figure}

Having determined the spatially converged results along spatial or mass direction, one can now undertake the full limit $\dr\rightarrow0,\dm\rightarrow0$ or $\dm\rightarrow0,\dr\rightarrow0$. The results of such double extrapolation are plotted in Figure \ref{fig:sfrall}, with the two limits coinciding almost perfectly. This is, of course, fully expected, as the order of taking the limits in $\dr$ and $\dm$ should not matter. The agreement, however, is not completely trivial, since extrapolations along both spatial and mass directions require fitting functions to noisy data; if the fits become unstable, the two limits would not agree. The actual agreement between them indicates the stability of the extrapolated values.

\section{Weak Numerical Convergence on Galactic Properties}
\label{sec:weak}

While finding the fully converged result is important, it is not practical to strive for strong convergence in production runs. Computing the solid line in Fig.\ \ref{fig:sfrall} requires producing 16 separate simulations (of which highest mass resolution runs dominate the total computational expense). The ultimate goal of a convergence study is, then, to achieve weak convergence - i.e., to find combinations of simulation parameters that allow to reproduce the fully converged result in simulations with finite spatial and mass resolutions. 

In addition, converging just on the global star formation history is not enough, since reproducing the global history does not guarantee that the detailed morphology of reionization or internal properties of galaxies are reproduced. Among these properties the most important for modeling reionization is star formation histories of individual galactic halos. Indeed, simulations with various mass and spatial resolution are expected to converge well on the distribution of sufficiently massive galactic halos, since modern N-body simulations are highly robust \citep{sims:kaa14}. Hence, as long as the mass resolution of a simulation is enough to resolve galaxies with halo masses $M_h\ga10^8\Msun$ that produce almost all ionizing photons \citep{ng:gk14}, one can attempt to construct a weakly converged model that will recover star formation histories of individual galactic halos and, hence, will reproduce the detailed reionization morphology of a fully converged solution.

\begin{figure}[t]
\includegraphics[width=\hsize]{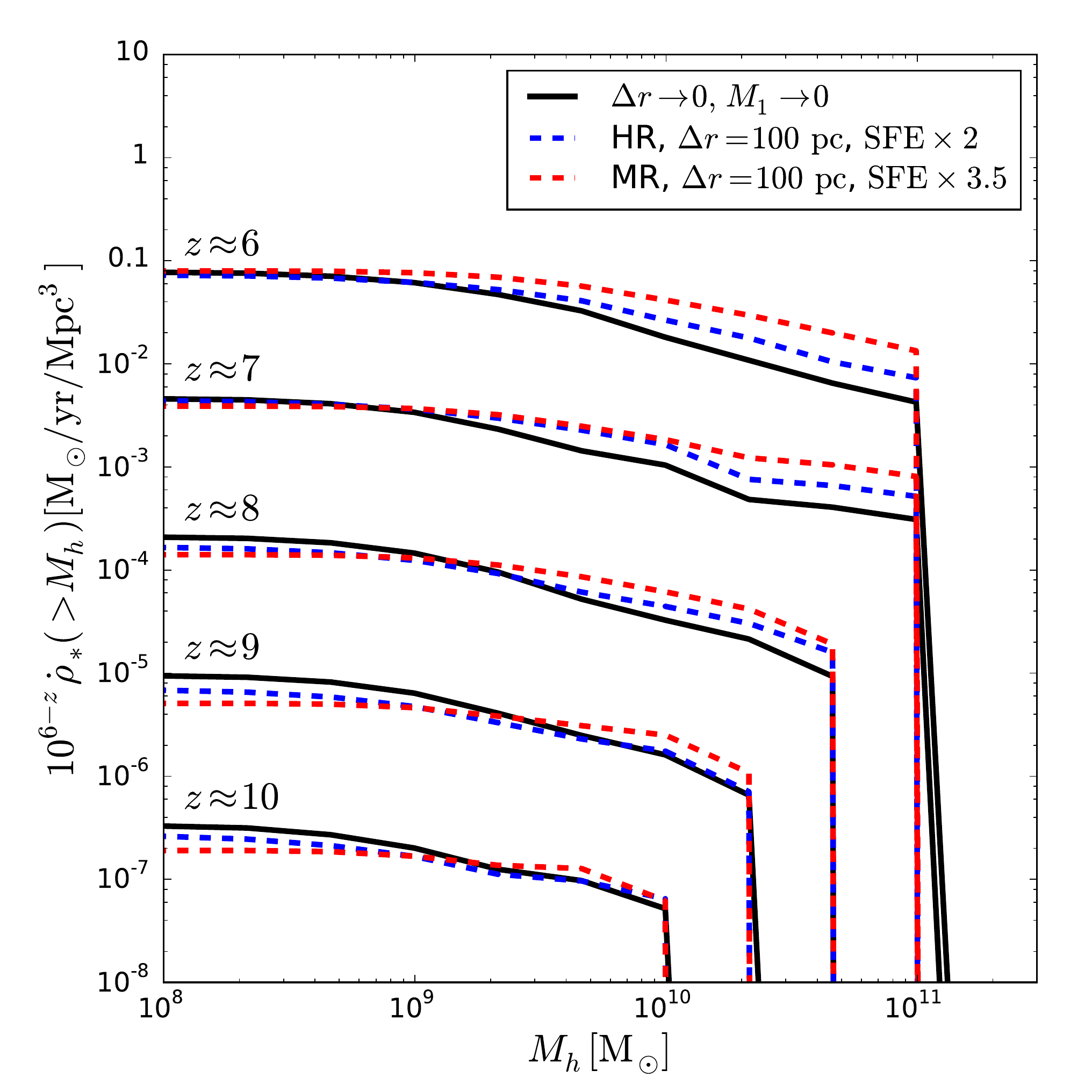}
\caption{Cumulative star formation rate density as a function of halo mass at several redshifts for the fully converged result (solid black lines) and typical finite resolution runs (red and blue dashed lines). Note that while the global SFR history is approximately matched by the finite resolution simulations (values of $\dot{\rho}_*$ at $M_h\rightarrow0$ as a function of redshift), the detailed star formation rates in halos of different masses are significantly off. Different redshifts are shifted vertically for clarity. \label{fig:rhosfraw}}
\end{figure}

Figure \ref{fig:rhosfraw} illustrates this point with an example of two finite resolution simulations. It shows a fraction of the global star formation rate density contributed to by galactic halos of a given mass, in a cumulative distribution, because in that case the value of $\dot{\rho}_*$ at $M_h\rightarrow0$ is simply the global star formation history (i.e. asymptotic values of black lines at $M_h\rightarrow0$ are simply values of solid lines in Fig.\ \ref{fig:sfrall} at these redshifts). 

Black solid lines show the fully converged ($\dm\rightarrow0$, $\dr\rightarrow0$) solutions, obtained with the methodology described in the previous section, and color lines give two typical finite resolution simulations, whose star formation efficiencies (SFE) are adjusted to match the global SFR history at $z=6$ (i.e. the agreement between all 3 lines for $M_h\rightarrow0$ at $z=6$ is by construction). While the global SFR histories for the latter are within $\sim25\%$ of the fully converged solution at all times, the overall shape of the distribution is not captured by the finite resolution simulations.

In order to fix this deviation, I introduce in the star formation recipe a multiplicative "weak convergence correction factor" $W$,
\[
  \dot{\rho}_* = W \times \dot{\rho}_*^{(\rm orig)},
\]
where $\dot{\rho}_*^{(\rm orig)}$ is the value of the star formation rate at a given location produces by the original, without the correcting factor, simulation recipe. Ideally, one would introduce a correction factor that depends on the halo mass - in that case it can simple be computed as the ratio of black and colored lines from Fig.\ \ref{fig:rhosfraw}. However, in practice that would require tracking halos as the simulation evolves, and would also introduce ambiguity during merger events. In addition, it would be a non-local modification of the star formation recipe. In order to avoid these purely technical, but still formidable complications, I introduce a weak convergence correction factor $W(n_{\rm H})$ that is a function of the total gas hydrogen density $n_{\rm H}$ only. It is, obviously, just a choice, and other forms of the conversion factor can be also designed. Hence, the weak convergence correction factor is not unique; even if two different forms of $W$ provide similar level of convergence for some particular galactic property, like $\dot{\rho}_*(>M_h)$, they may result in different degree of convergence for some other galactic properties. It is unclear, however, how one would approach finding an "optimal" convergence factor.

\begin{figure}[t]
\includegraphics[width=\hsize]{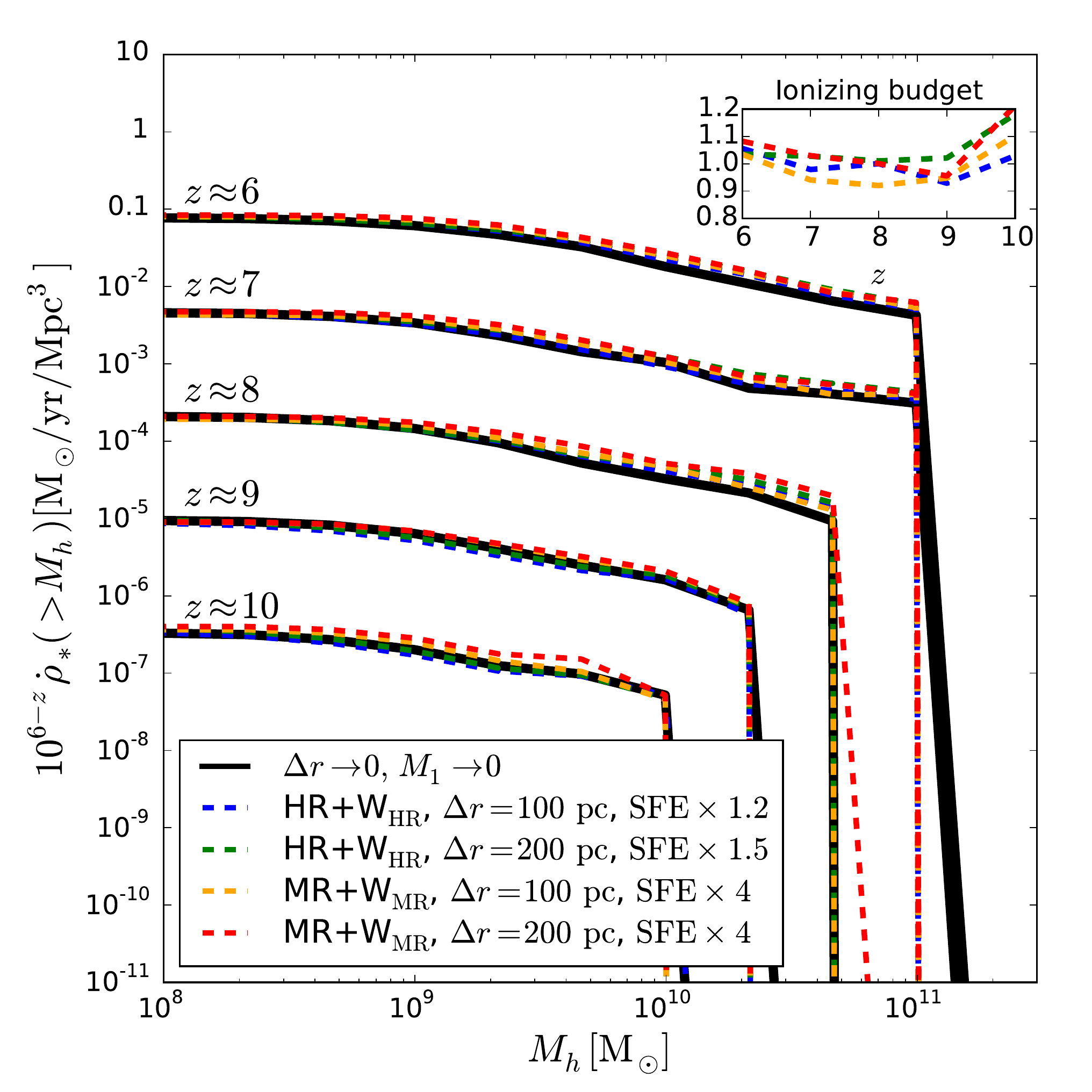}
\caption{Cumulative star formation rate density as a function of halo mass at several redshifts for the fully converged result (solid black lines) and four weakly converged finite resolution simulations. With the weak convergence correction the finite resolution simulations recover the whole shape of $\dot{\rho}_*(>M_h)$ at all halo masses with an acceptable ($\sim20\%$) precision. The insert shows the ratios of $\dot{\rho}_*(M_h\rightarrow0)$ from four weakly converged simulations and the fully converged solution, which is the precision with which weakly converged simulations recover the total star formation rate and, hence, the global ionizing budget.\label{fig:rhosfweak}}
\end{figure}

Hence, to be specific, I choose the following form for $W(n_{\rm H})$,
\begin{equation}
  W(n_{\rm H}|q_<,q_>,n_c) = \frac{q_< + q_>(n_{\rm H}/n_c)}{1+(n_{\rm H}/n_c)},
  \label{eq:wc}
\end{equation}
where $q_<$, $q_>$, and $n_c$ are parameters. Extensive parameter search results in the following weak convergence correction factors for ``High Resolution'' ($\dm=9.2\times10^5\Msun$) runs,
\begin{equation}
  W_{\rm HR}(n_H) \equiv  W(n_{\rm H}|3,1,10\dim{cm}^{-3}),
  \label{eq:wchr}
\end{equation}
and ``Medium Resolution'' ($\dm=7.4\times10^6\Msun$) runs,
\begin{equation}
  W_{\rm MR}(n_H) \equiv  W(n_{\rm H}|3,0.3,10\dim{cm}^{-3}).
  \label{eq:wcmr}
\end{equation}
Both factors enhance star formation in low mass halos; the ``Medium Resolution'' form also reduces star formation rate in highest density regions, to force a larger variation in $\dot{\rho}_*(>M_h)$ with the halo mass. The actual values of the parameters $q_<$, $q_>$, and $n_c$ are rather robust; $n_c$ can be varied between $3\dim{cm}^{-3}$ and $30\dim{cm}^{-3}$ with little change, and $q_<$ and $q_>$ can also be varied by up to 50\%.

Figure \ref{fig:rhosfweak} now shows $\dot{\rho}_*(>M_h)$ vs $M_h$ for four weakly converged simulations, ``High Resolution'' and ``Medium Resolution'' runs with $100\dim{pc}$ and $200\dim{pc}$ (with the appropriate adjustment in the SFE\footnote{Remember, that the SFE (quantified by the gas depletion time $\tsf$) is a free parameter of these simulations, to be fixed by matching some data -  for example, the observed galaxy UV luminosity functions. Hence, in this paper I am primary concerned with the relative star formation rates in various runs, with the absolute scale to be determined later, after the fully converged solution is compared to the observations.}). The overall shapes are now reproduced at an acceptable level of precision, and the global star formation history (aka the total ionizing budget) are also recovered to within 10\% for $100\dim{pc}$ runs and within 20\% for  $200\dim{pc}$ simulations.

\begin{figure}[t]
\includegraphics[width=\hsize]{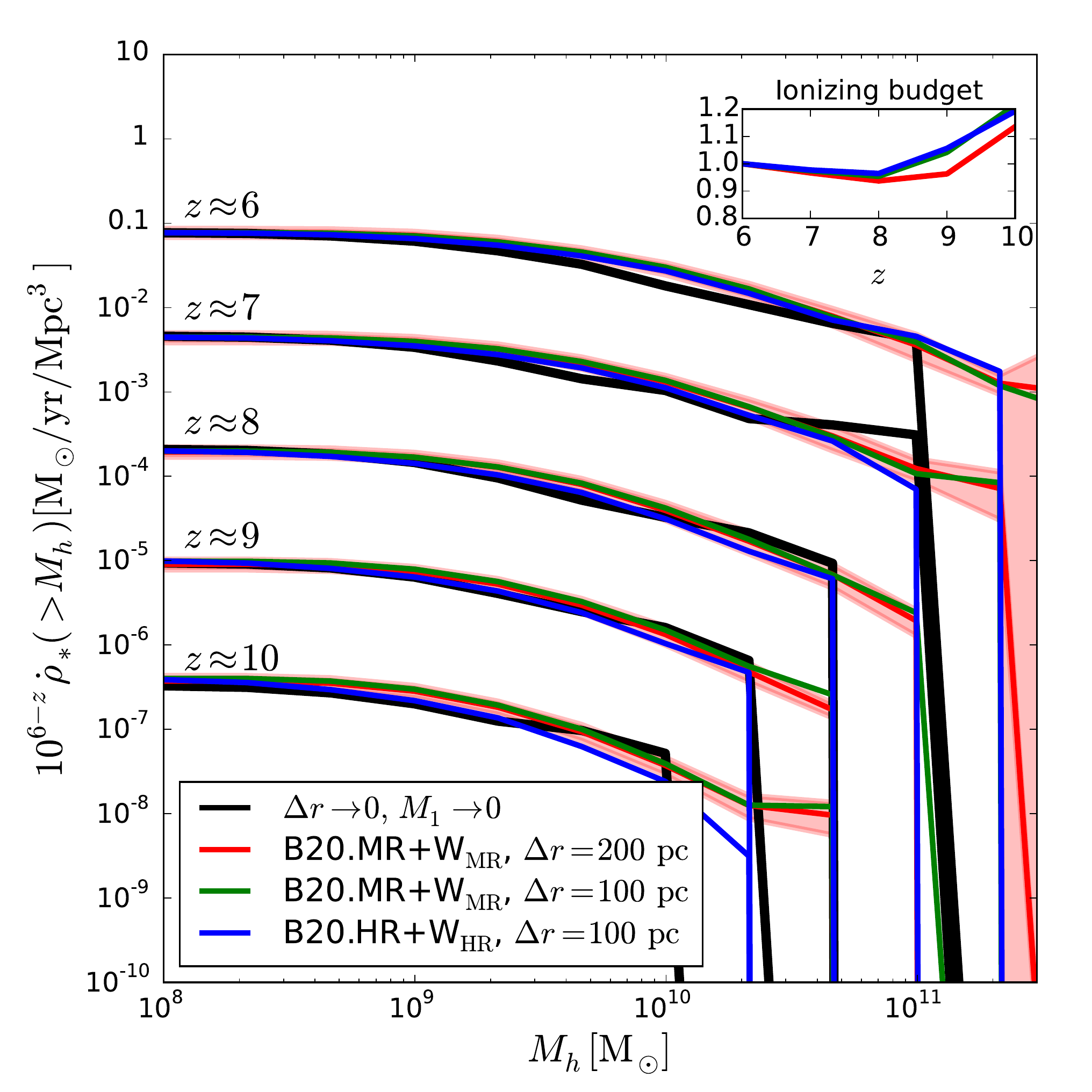}
\caption{Cumulative star formation rate density as a function of halo mass at several redshifts for the fully converged result (solid black lines) and three weakly converged finite resolution simulations in $20h^{-1}\dim{Mpc}$ boxes (two ``Medium Resolution'' sets with $100\dim{pc}$ and $200\dim{pc}$ spatial resolution and a single ``High Resolution'' run with $100\dim{pc}$ spatial resolution). A semi-transparent band shows the error in the mean computed from 6 random realizations.\label{fig:rhosfb20}}
\end{figure}

While my primary focus in this paper are spatial and mass resolution, the simulation box size is also an important numerical parameter. In Figure \ref{fig:rhosfb20} I show weak convergence for several simulations in $20h^{-1}\dim{Mpc}$ boxes (as compared to $10h^{-1}\dim{Mpc}$ for all the runs presented above): two ``Medium Resolution'' ($512^3$) simulation sets with $100\dim{pc}$ and $200\dim{pc}$ spatial resolution and a single ``High Resolution'' ($1024^3$) run with $100\dim{pc}$ spatial resolution. Each ``Medium Resolution'' box was simulated 6 times in independent realizations of initial conditions \citep[similar to runs B20.sf1.uv2.bw10 from][]{ng:g14}, and the figure shows the average over all 6 realizations and its error (for the $200\dim{pc}$ case; the error for the $100\dim{pc}$ is essentially identical). Simulations with twice larger box sizes achieve a similar level of weak convergence irrespectively of their spatial or mass resolution, illustrating the approximate independence of the accuracy of weak convergence from the box size.

Since weak convergence correction factors are tuned to recover the fully converged solution specifically for $\dot{\rho}_*(>M_h)$, independent tests of the quality of these modified runs can be obtained by comparing numeric convergence for other galactic properties - since the factors $W$ are fully fixed in these runs, all other galactic properties are uniquely predicted in the simulations.

\begin{figure*}[th]
\includegraphics[width=0.33\hsize]{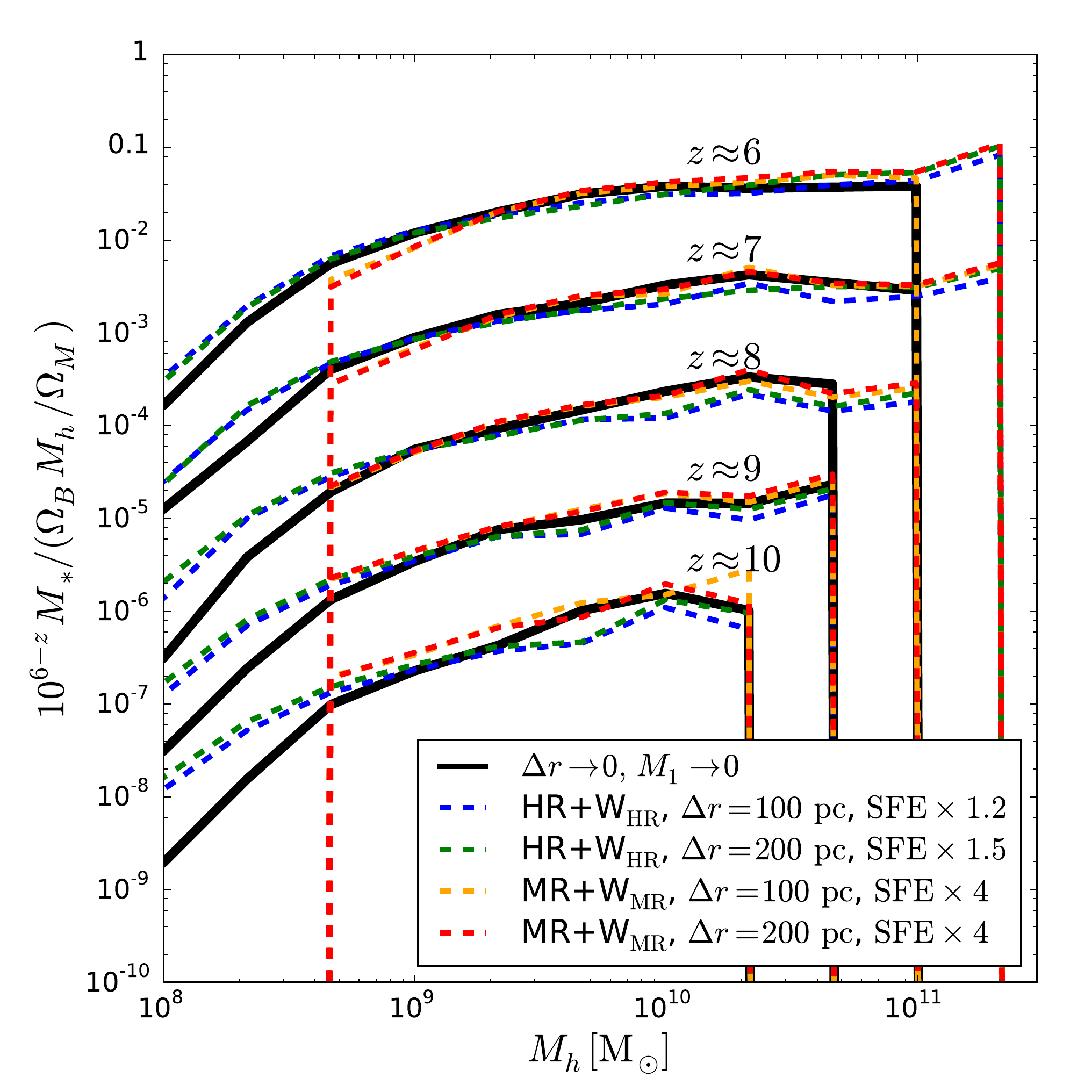}%
\includegraphics[width=0.33\hsize]{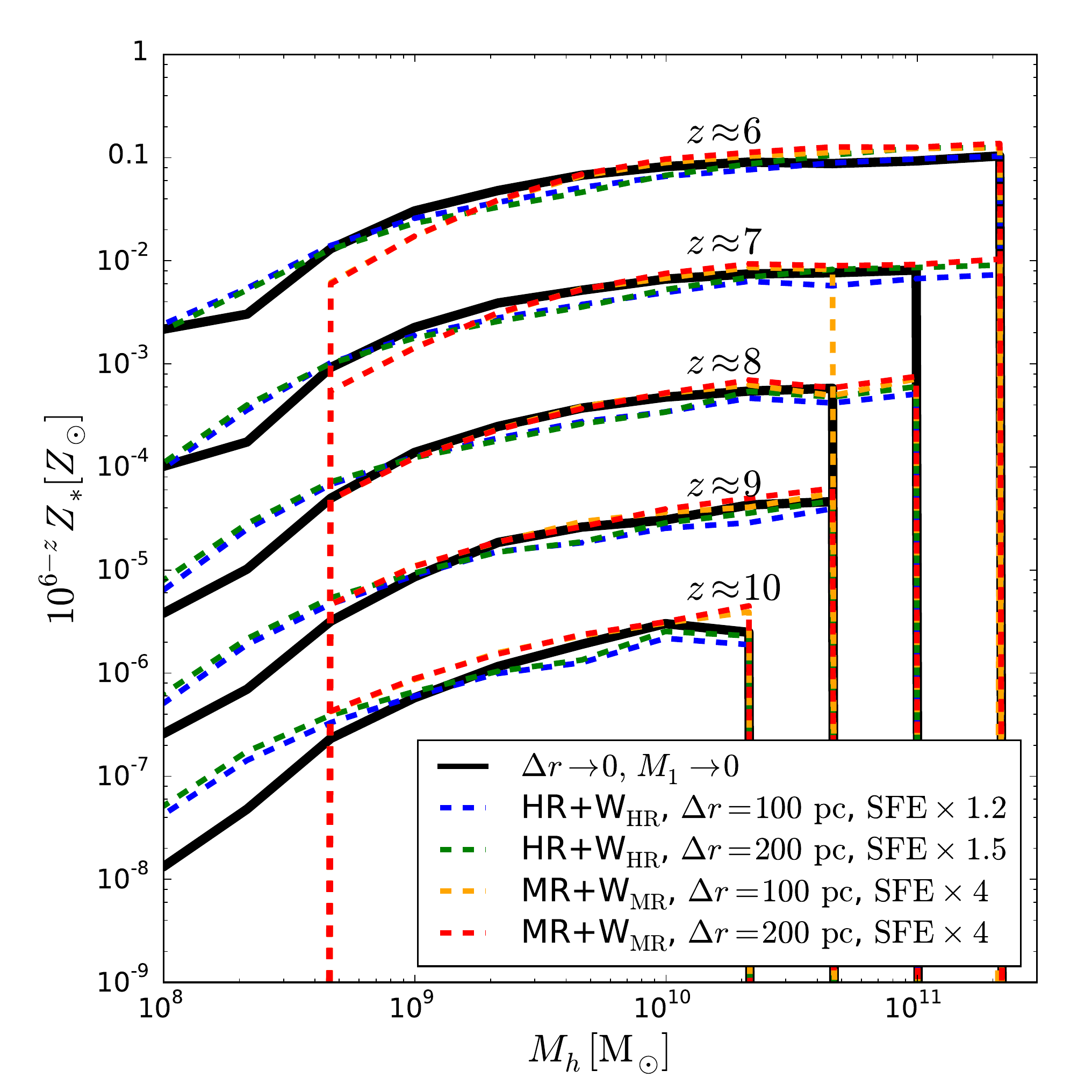}%
\includegraphics[width=0.33\hsize]{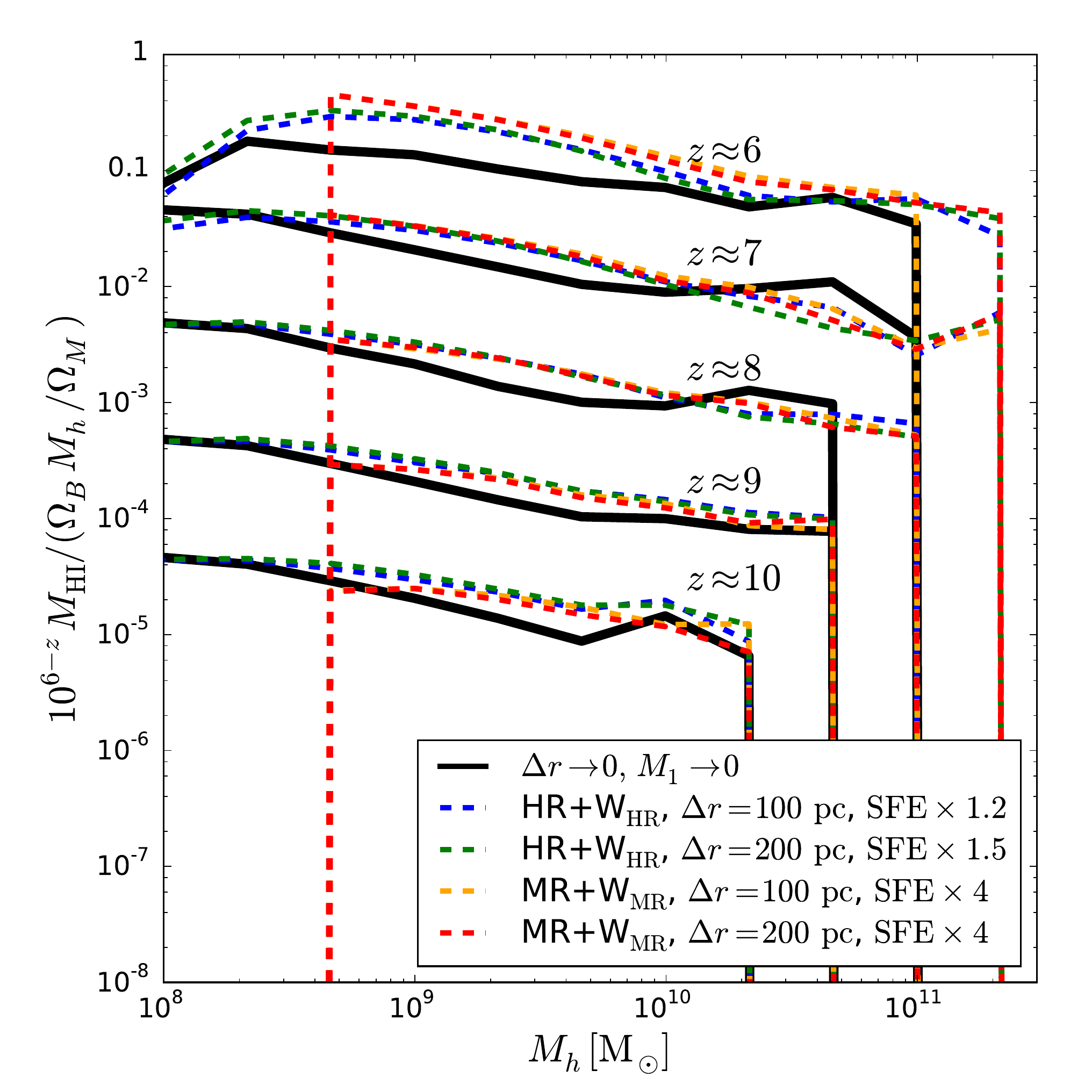}%
\caption{Tests of numerical convergence for $M_*$ vs $M_H$, $Z_*$ vs $M_h$, and $M_\HI$ vs $M_h$ for the weakly converged runs shown in Fig.\ \ref{fig:rhosfweak}. A good level of convergence for $M_*$ and $Z_*$ is not that surprising, since converging on full star formation histories of individual galactic halos effectively guarantees that. Agreement with atomic gas fractions - a truly independent test - comes out significantly worse.\label{fig:otherweak}}
\end{figure*}

In Figure \ref{fig:otherweak} I show three other galactic properties for the fully converged solution and four weakly converged simulations discussed above. A decent level of convergence for stellar masses and metallicities is not surprising, since converging on full star formation histories of individual galactic halos effectively guarantees that most of stellar properties are well converged as well. However, a given star formation history does not constraint many other galactic properties, for example, atomic hydrogen fraction. As the right panel of Fig.\ \ref{fig:otherweak}, this quantity is converged much worse, at the level of a factor of 2 only. Hence, the weakly converged simulations should be used with care when making predictions for, for example, 21 cm emission - they will predict correctly the emission from the large-scale distribution of ionized bubbles (since star formation rates and, hence, ionizing luminosities of individual galactic halos are recovered), but will overestimate the additional contribution from galaxies themselves, which can be important at $\sim25-50\%$ level \citep{ng:kg15b}.

An alternative, and likely a better approach (suggested by the referee) could be to modify the actual gas density that enters star formation model and radiative transfer solver (but not the hydro solver, which needs to remain strictly conservative). It may be possible then with a single factor to achieve weak convergence in several simulated quantities, such as star formation rates, metallicities, $\HI$ masses, and even the interstellar radiation field. Finding such a universal correction will be, though, a significant effort.

\section{Conclusions}
\label{sec:con}

Because numerical simulations offer the most accurate and realistic theoretical models of reionization at present, evaluating their precision is an important theoretical task. Results from computer simulations can be considered as solutions to actual physical equations only if they are numerically converged. In practice, however, it is often exceptionally difficult or even plainly impossible to reach "strong" convergence - i.e., the effective independence of simulation results of spatial and mass resolution (see eq.\ \ref{eq:sconv}) - and one has to seek instead "weak" convergence. With weak convergence the simulated physical model at finite spatial and mass resolution is adjusted as the resolution changes (eq.\ \ref{eq:wconv}). One can consider such an adjustment as tuning a finite resolution simulation to reproduce the results of the fully converged solution, a completely legitimate and sensible approach to take.

In this paper I show that simulations of reionization performed under the Cosmic Reionization On Computers (CROC) project do converge in space and mass, albeit slower than one hoped for. A fully converged solution can be obtained at a level of about 20\% precision, and which is also independent of the adopted model for molecular hydrogen formation. While this can be considered an important achievement, it is useless in practice, since populating the grid of various resolution values  for production grade simulation, needed for accurate extrapolation to formally infinite resolution, would require an insane amount of computational time.

In order to make progress in the interim, before such large computer allocations become possible, I introduce a weak convergence correction factor in the star formation recipe, which allows to approximate the fully converged solution with finite resolution simulations. The accuracy of weakly converged simulations approaches a comparable, $\sim20\%$ level of precision for star formation histories of individual galactic halos and other galactic properties that are directly related to star formation rates, like stellar masses and metallicities. Yet other properties of model galaxies, for example, their $\HI$ masses, are recovered in the weakly converged runs to only a factor of two.

Overall, the weakly converged solutions will serve as reasonable interim theoretical counterparts for the first observational data from JWST, until more expensive, better converged simulations become available.

\acknowledgements

I am grateful to the anonymous referee for insightful comments and an important suggestion for the future improvement of the current work.
Simulations used in this work have been performed on the Joint Fermilab - KICP cluster ``Fulla'' at Fermilab, on the University of Chicago Research Computing Center cluster ``Midway'', and on National Energy Research Supercomputing Center (NERSC) supercomputers ``Hopper'' and ``Edison''.

\appendix

\section{Implementation of Fixed Proper Resolution in the ART Code}
\label{app:resol}

Grid codes (including AMR) commonly maintain spatial resolution fixed in comoving coordinates, while SPH codes often hold resolution fixed in proper coordinates. The latter is a desirable property, since most subgrid physical models work well over a limited range of spatial scales. Hence, in a grid simulation the proper resolution may, with time, redshift away from the preferred range. A simple, approximate implementation of fixed proper resolution in an AMR code has been recently presented by \citet{sims:rta14}. In their scheme a new refinement level is added every time the proper resolution redshifts too much away from a preset value. 

A \citet{sims:rta14} approach is very simple, but it has one undesirable feature: as a new AMR level is added, the actual spatial resolution of the simulation jumps by a factor of 2. Since the new level was held back artificially, a large fraction of the parent level would have been refined were that level active; as the new level is released, a significant fraction of cells on the parent level gets refined to the new level on a very short time scale (free-fall time-scale of the new level). Because, as we have shown above, CROC simulations do \emph{not} achieve strong convergence in spatial resolution, such a jump in resolution creates a comparable jump in the global star formation rate. Such a ``universe-quake'' behavior is, obviously, totally unphysical and even violates causality.  

In order to avoid unphysical behavior, a different scheme is implemented in the ART code that includes several components.
\begin{description}
\item[Slow down, but not suppress refinement below $\dr$.] No refinement level is held back artificially, but the refinement criterion is modified so that on any level $L$ below the fixed proper resolution $\dr$ the dark matter and/or gas mass in a cell required for refinement to a higher level $L+1$ is increased by a factor $\left(\Delta x_L/\dr\right)^3$, so that only the densest cells would refine below the fixed proper resolution. 
\item[Smooth the source term for the Poisson equation.] The total density that sources the Poisson equation is smoothed on a scale comparable to $\dr$. Namely, for a given value of $\dr$ two neighboring refinement levels $L_d$ and $L_u=L_d+1$ are found such that $\Delta x_{L_u} \leq \dr$ and $\Delta x_{L_d} > \dr$. Then the density on level $L_u$ that is used in the Poisson equation is computed as
\[
  \tilde\rho_{L_u} = (1-w)\rho_{L_u} + w\rho_{L_d}
\]
with $w = \log_2(\dr/\Delta x_{L_u})$ and $\rho_L$ being the actual total density on the refinement level $L$. On all levels $L\leq L_d$ the actual total density $\rho_L$ is used in the Poission equation, and for all levels $L>L_u$ the density of the parent cell at level $L_u$ is used in the Poission equation. This procedure is not precisely a convolution with a given window function, but it is (a) linear and (b) gives $\tilde\rho_L = \rho_L$ for $\dr=\Delta x_L$, so it serves as a smoothing procedure for $\tilde\rho$.
\item[Maintain pressure floor consistently with gravity.] The pressure floor that prevents numerical fragmentation in the gas is maintained so that the local Jeans length is always resolved with at least $4\dr$.
\end{description}

\begin{figure*}[th]
  \includegraphics[width=0.5\hsize]{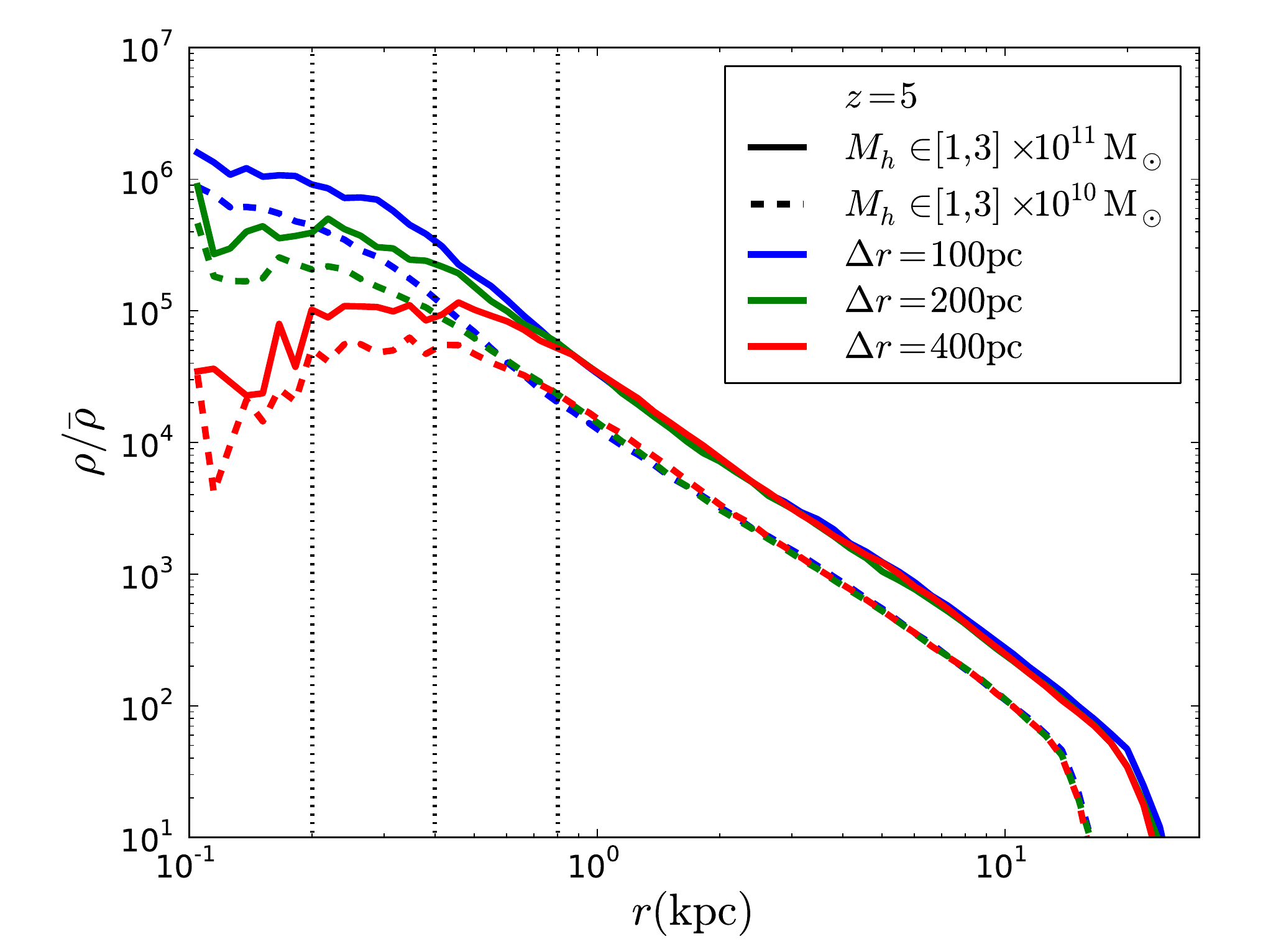}%
  \includegraphics[width=0.5\hsize]{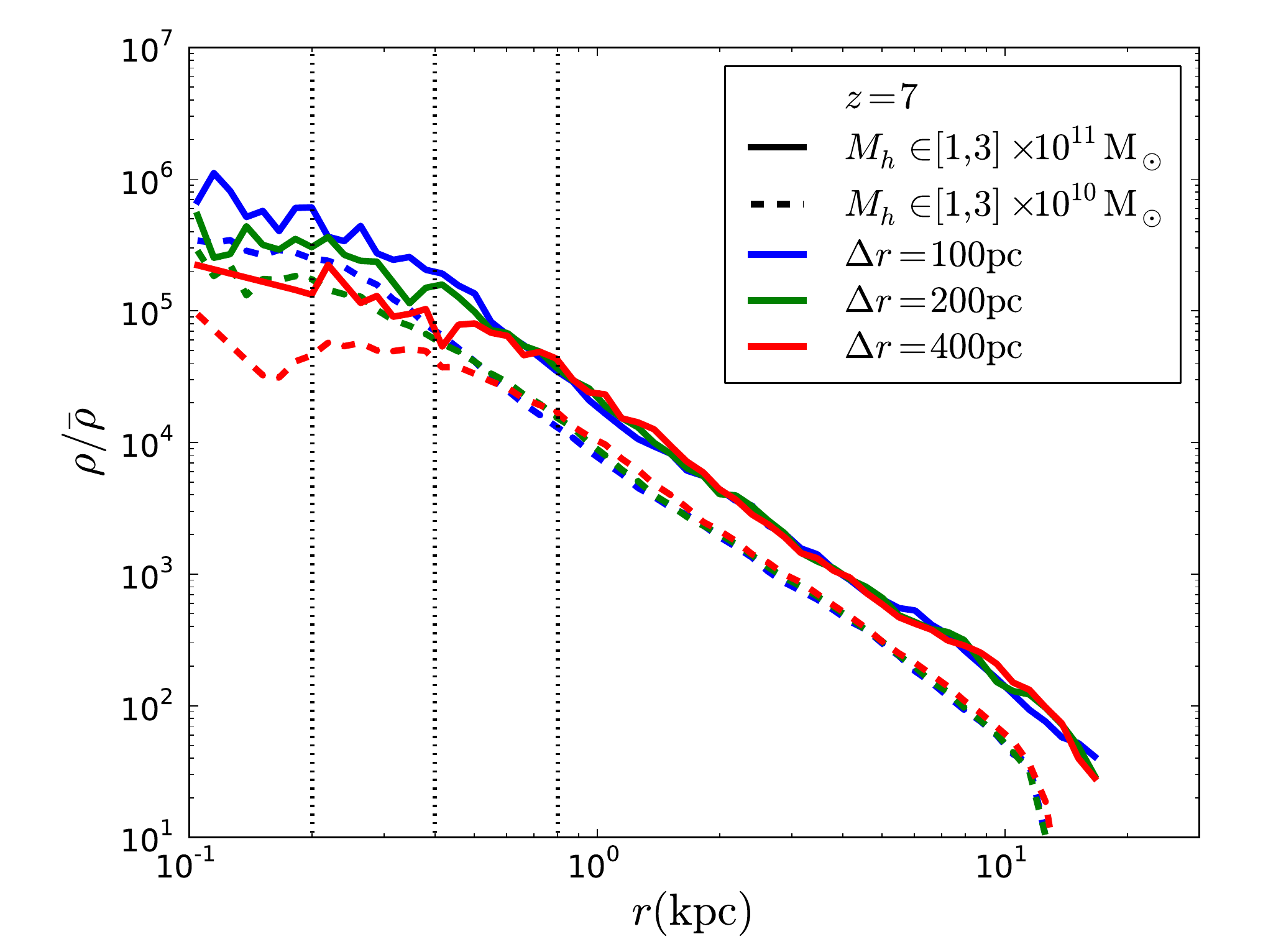}
  \caption{Average dark matter density profiles for halos (in proper distance coordinates) in two mass bins and for three spatial resolutions in B10.HR simulations. Vertical dotted lines show radii of $200\dim{pc}$, $400\dim{pc}$, and $800\dim{pc}$ - a run with $\dr=200\dim{pc}$ has the real spatial resolution about twice worse. Two panels show $z=5$ and $z=7$ to demonstrate that the spatial resolution is indeed reasonably redshift independent. \label{fig:prof}}
\end{figure*}

Average density profiles in halo mass bins for two test simulations with $\dr=100\dim{pc}$, $\dr=200\dim{pc}$, and $\dr=400\dim{pc}$ are shown in Figure \ref{fig:prof}. As one can see, the real resolution of the simulations is about $2\times\dr$, as could be expected for a grid code, and the resolution is essentially the same at two different redshifts, i.e.\ it remains constant in proper coordinates.

\section{Molecular Hydrogen Formation Model}
\label{app:h2mod}

The target spatial resolution for CROC simulations is $100-200\dim{pc}$, and a detailed chemical model for molecular hydrogen formation would not work at such resolution \citep{ng:gk11}. Instead, it is appropriate to use the fitting formulas of \citet{ng:gd14} (GD14) that are specifically designed to work on scales of several hundred parsecs.

\begin{figure}[b]
  \centerline{\includegraphics[width=0.5\hsize]{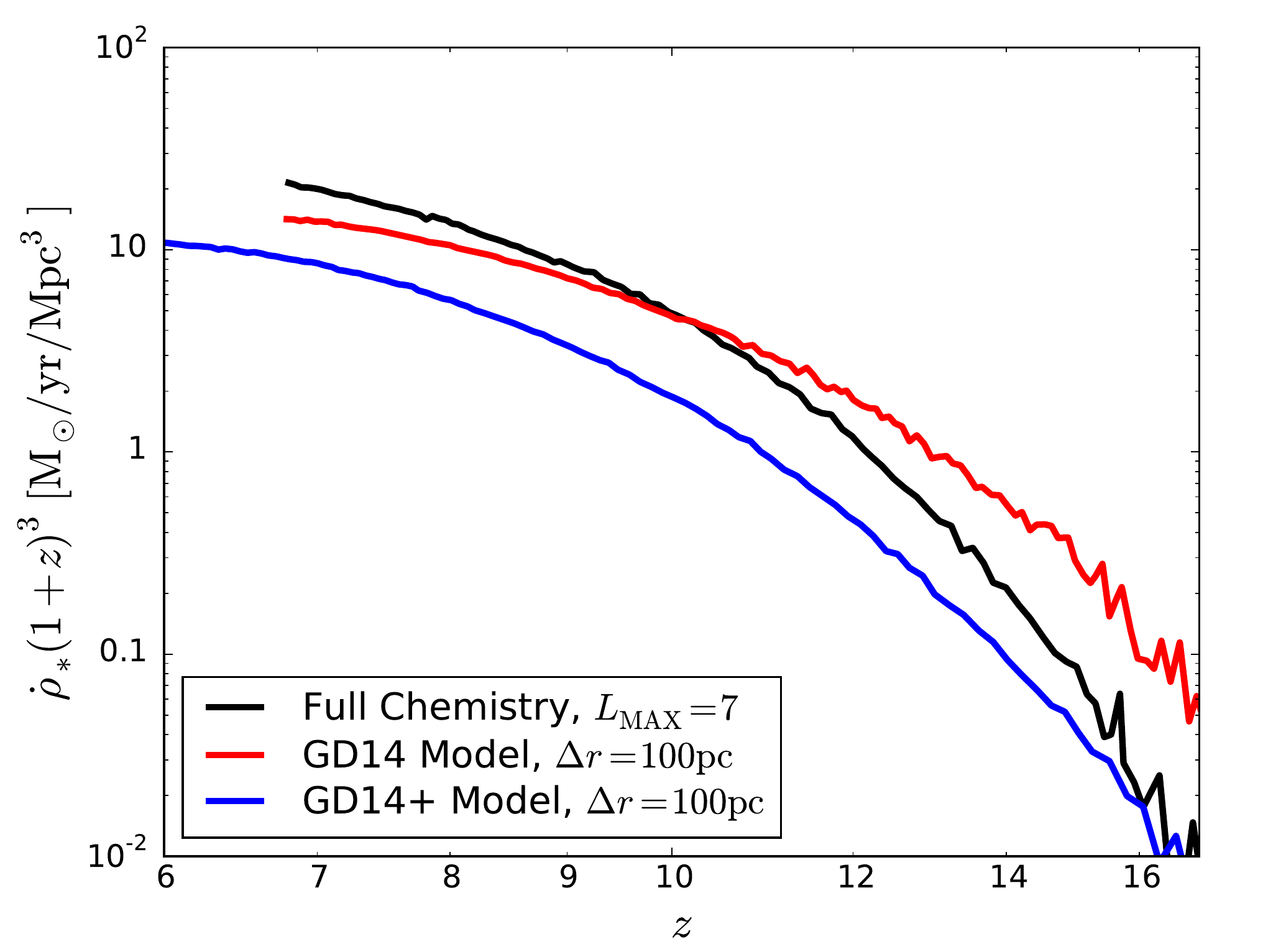}}
  \caption{Global star formation histories of several test ``HR'' ($\dm=9.2\times10^5\Msun$) simulations. A black line shows the reference ``Full Chemistry'' HR run with 7 levels of refinement (run \bfull{10}{HR}{L7}). A red line presents the simulation that implements the original GD14 fits as a molecular hydrogen formation model with $100\dim{pc}$ resolution (run \bgd14{10}{HR}{}{100}). Large excess of the GD14 model over the reference model at $z\ga11$ is a manifestation of the  failure of GD14 fits at early times. A blue line shows the corrected ``GD14+'' model that eliminates the unphysical behavior (run \bgd14{10}{HR}{+}{100}).\label{fig:sfrnma}}
\end{figure}

There exists, however, one serious limitation of the GD14 fitting formulas: they were calibrated on lower redshift simulations, and do not account for the fact that molecular hydrogen formation time may be long, even longer than the age of the universe. At $z\sim3$ this is not an issue, since the age of the universe is $2\dim{Gyr}$, comparable to the molecular hydrogen depletion time. At high redshifts, $z\ga10$, the effect of the finite age of the universe is, however, substantial. This is illustrated in Figure \ref{fig:sfrnma}, which compares a reference, ``Full Chemistry'' simulation (run \bfull{10}{HR}{L7}) with a test simulation with the same mass resolution (``HR''), but implementing the original GD14 fits as a model for molecular hydrogen abundance (run \bgd14{10}{HR}{}{100}). At $z \ga11$, GD14 fits predict star formation rates (and, hence, molecular hydrogen abundance) in excess of the ``Full Chemistry'' model. Since the ``Full Chemistry'' solves the actual time-dependent chemical network of reactions, it properly accounts for the time available for molecular hydrogen formation in low dust abundance or low density environments. Hence, the high star formation rate returned by the GD14 test simulation is unphysical.

In order to correct this unphysical behavior, I modify the original GD14 fits as follows. The molecular hydrogen fraction in the gas is computed as 
\begin{equation}
  f_\H2 = \frac{\tilde{R}}{1+\tilde{R}},
  \label{eq:fit1}
\end{equation}
where $\tilde{R}$ is the ratio of the molecular to atomic gas, and is parametrized as
\begin{equation}
  \tilde{R} = \frac{q}{1+q}R,
  \label{eq:fit2}
\end{equation}
where $R$ is given by unmodified GD14 fits, their Eqs.\ (8-10). In the limit $q\rightarrow\infty$, this modified, ``GD14+'' fit reduces to the original GD14 fitting formula.

There is no rigorous way to compute the correction factor $q$ (short of running a ``Full Chemistry'' simulation), so it has to be implemented with a model. The average rate of formation of molecular hydrogen in a simulation cell is 
\[
  \frac{\dot{n}_\H2}{n_\H2} =   3.5\times10^{-17}\dim{cm$^3$/s} \times D_{\rm MW} n_{\rm H} C_d ,
\]
where $D_{\rm MW}$ is the dust-to-gas ratio in the Milky Way units (i.e., in the solar neighborhood $D_{\rm MW}=1$) and $C_d$ is the clumping factor that accounts for the numerically unresolved density stricture on small scales \citep[since formation of $\H2$ is a two-body process - fuller details, including equations, are given in][]{ng:gk11,ng:gd14}. A plausible ansatz for the suppression factor $q$ is  
\[
  q \sim \frac{\dot{n}_\H2}{n_\H2} \tau_c,
\]
where $\tau_c$ is some time scale. From general consideration, it is hard to guess what it should be. For example, it can be proportional to the age of the universe, in which case $q$ scales as $D n_{\rm H}$; it can also be proportional to the free-fall time $t_{\rm ff}$, in which case $q$ scales as $D n_{\rm H}^{1/2}$; it can also scale with density as the cooling time, $\tau_c \propto 1/n_{\rm H}$, in which case $q$ scales as $D$ with no density dependence. 

\begin{figure}[t]
  \includegraphics[width=0.5\hsize]{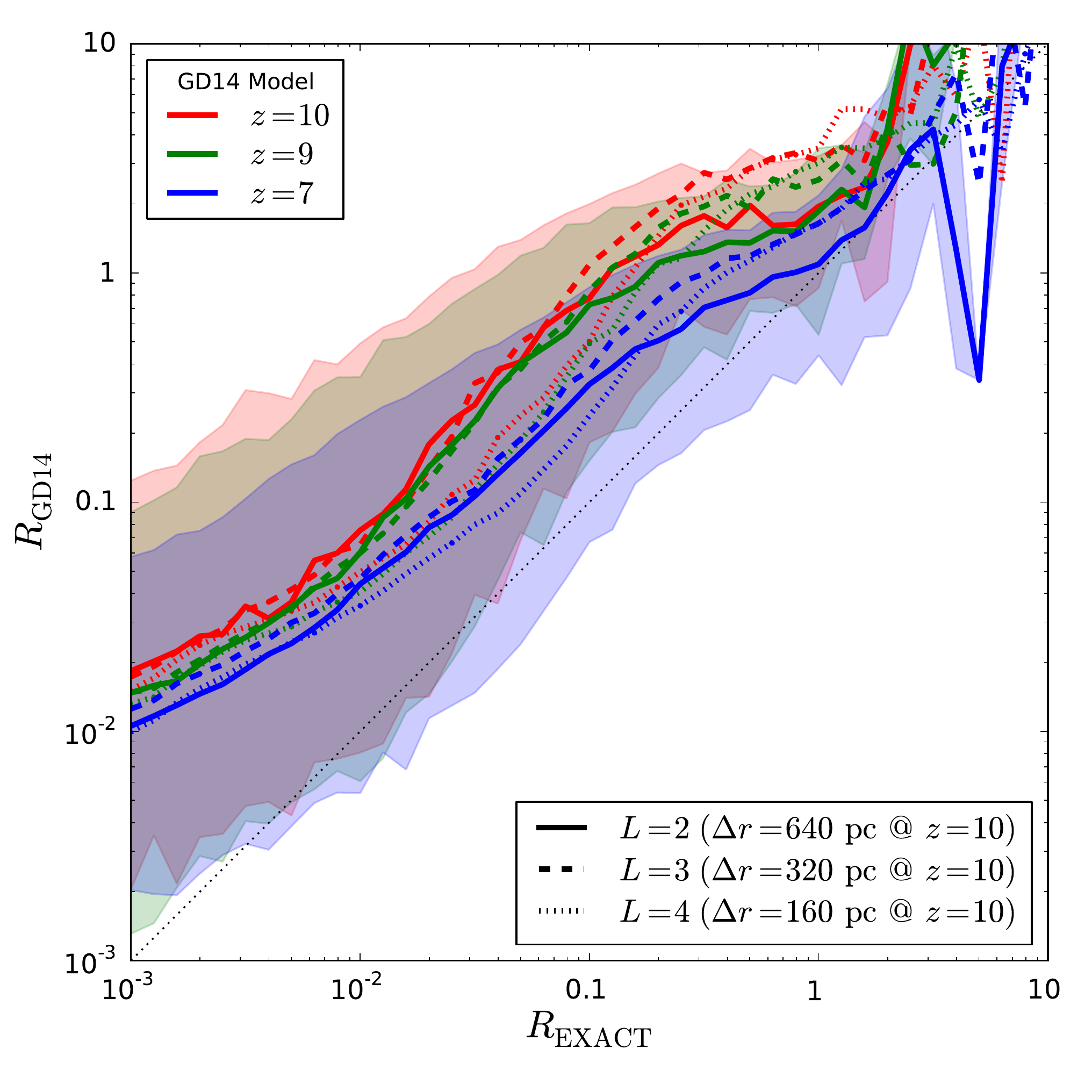}%
  \includegraphics[width=0.5\hsize]{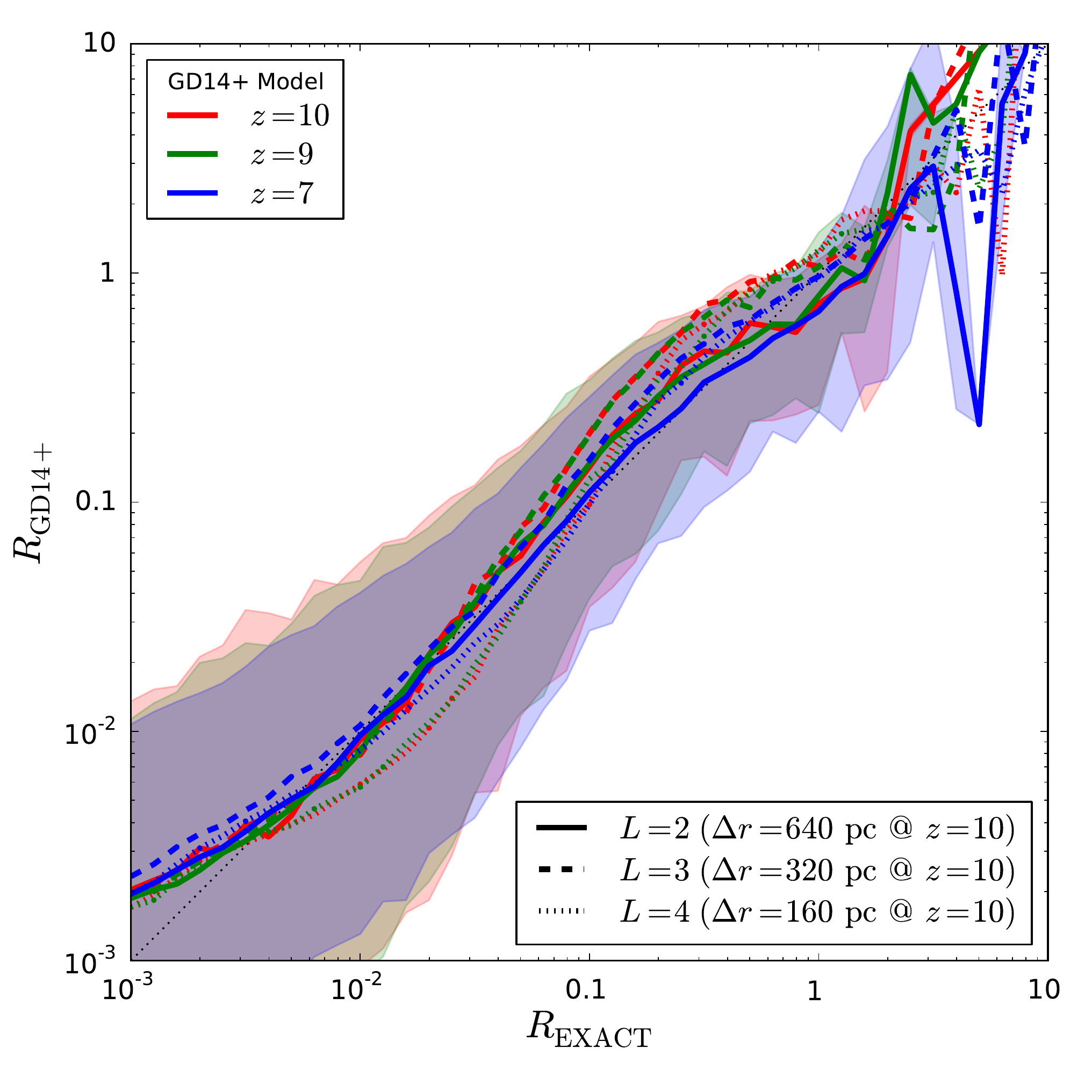}
  \caption{Comparison of the actually simulated molecular-to-atomic gas ratio $R\equiv\rho_\H2/\rho_\HI$ versus the one computed using GD14 fits at two redshifts, $z=10$ (red) and $z=7$ (blue) and three different resolutions. Lines show median values, while semi-translucent bands show 10\%-90\% percentile range. Two panels show the original GD14 fits and the modified GD14+ model from Equation (\protect\ref{eq:qmod}).\label{fig:gd14}}
\end{figure}

In order to explore this choices, I use the "Full Chemistry" run \bfull{10}{HR}{L7} and apply GD14 fitting formula to the actual simulated data, averaged over less refined levels (thus, mimicking lower resolution). Such a comparison is shown in the left panel of Figure \ref{fig:gd14} (it can be directly compared to Fig.\ 5 of GD14). The agreement between the actual calculation and the GD14 fitting formula is not nearly as good as in the GD14 paper, and that is the manifestation of the finite value of $\tau_c$. I explored all three possible ansatzes for the $q$ factor discussed above ($q\propto D n_{\rm H}^\alpha$ with $\alpha=0$, $1/2$, and $1$), and the best result is obtained for $\alpha=0$ case.  In that case the factor $q$ can be parametrized as
\begin{equation}
  q = \left(\frac{D}{D_c}\right)^{1/2},
  \label{eq:qmod}
\end{equation}
with
\begin{equation}
  D_c = 0.05\left(\frac{1+z}{10}\right)^3.
  \label{eq:qdz}
\end{equation}
Comparison between the exact calculation and the fit for this case is shown in the right panel of Fig.\ \ref{fig:gd14}. This particular ansatz has several desirable features: for example, the redshift dependence of $D_c$ eliminates the redshift dependence in the right panel of Fig.\ \ref{fig:gd14}. In addition, at low redshifts ($z\la5$) the factor $D_c$ becomes very small, and the modified model reduces to the original GD14 fits.

I, therefore, adopt ansatz (\ref{eq:qmod}) as my molecular hydrogen formation model, and label it "GD14+" to underscore a small but important correction to the original GD14 fit. A corresponding test simulation (run \bgd14{10}{HR}{+}{100}) is shown in Fig.\ \ref{fig:sfrnma} with a blue line. As one can see, the unphysical behavior is now eliminated.

\bibliographystyle{apj}
\bibliography{ng-bibs/self,ng-bibs/ism,ng-bibs/misc,ng-bibs/sims,ng-bibs/sfr,ng-bibs/rei,ng-bibs/dsh,ng-bibs/qlf,ng-bibs/gals,ng-bibs/igm,ng-bibs/reisam}

\end{document}